\documentclass[fleqn,usenatbib]{mnras}

\usepackage{newtxtext,newtxmath}
\usepackage[T1]{fontenc}

\DeclareRobustCommand{\VAN}[3]{#2}
\let\VANthebibliography\thebibliography
\def\thebibliography{\DeclareRobustCommand{\VAN}[3]{##3}\VANthebibliography}

\usepackage{graphicx}	
\usepackage{booktabs} 
\usepackage{array}
\usepackage{ulem}
\newcommand{\feh}{$\rm [Fe/H]$}
\newcommand{\teff}{$T_{\rm eff}$}
\newcommand{\M}{$M_{\odot}$}

\usepackage{xcolor}
\definecolor{orcidgreen}{RGB}{166, 206, 57}
\usepackage{fontawesome5}
\usepackage{hyperref}
\newcommand{\orcid}[1]{%
  \href{https://orcid.org/#1}{\textcolor{orcidgreen}{\faOrcid}}%
}

\title[The IMF of low-mass stars]{The Initial mass function of field stars with mass $\leq$1 $M_{\odot}$ varies with metallicity}

\author[Qiu et al.]{
Dan Qiu,$^{1,2,3}$\orcid{0000-0002-8280-4808},
 Chao Liu,$^{2,3,4,5}$\thanks{E-mail: liuchao@nao.cas.cn},
 Jennifer A. Johnson,$^{6}$\orcid{0000-0001-7258-1834}
 Jiadong Li,$^{7}$\orcid{0000-0002-3651-5482}
 Bo Zhang$^{2}$\orcid{0000-0002-6434-7201}
\\
$^1$Kavli Institute for Astronomy and Astrophysics, Peking University, Beijing 100871, People’s Republic of China;\\
$^2$ Key Laboratory of Space Astronomy and Technology, National Astronomical Observatories, CAS, Beijing 100101, People’s Republic of China; \\
$^3$ University of Chinese Academy of Sciences, Beijing 100049, People’s Republic of China;\\
$^4$Institute for Frontiers in Astronomy and Astrophysics, Beijing Normal University, Beijing 100875, People’s Republic of China;\\ $^{5}$Zhejiang Lab, Hangzhou, Zhejiang 311121, People’s Republic of China;\\
$^6$ Department of Astronomy, The Ohio State University, 140 W. 18th Ave., Columbus, OH 43210, USA;\\
$^7$Max-Planck-Institut für Astronomie, Königstuhl 17, D-69117 Heidelberg, Germany;
}

\date{Accepted XXX. Received YYY; in original form ZZZ}

\pubyear{2015}

\begin{document}
\label{firstpage}
\pagerange{\pageref{firstpage}--\pageref{lastpage}}
\maketitle

\begin{abstract}

We investigated a volume-limited sample of LAMOST main-sequence stars with masses from 0.25 to 1 \M\ and distances of 150-350 pc to explore how the stellar initial mass function (IMF) varies with metelliaicty. 
We corrected the spectroscopic selection function by comparing the stellar number densities with the photometric ones at the same colour and magnitude. From these corrected number density distributions, we derived IMFs for each metallicity sub-samples. Fitting a broken power-law function in each IMF with a fixed break point at 0.525 \M, we found the power-law indices increase with
[Fe/H] for both mass regimes: $\alpha_1$ (mass $\leq$0.525\M) rises from 0.54$\pm$0.21 to 1.40$\pm$0.07 and $\alpha_2$ (mass$>$0.525\M) grows from 1.40$\pm$0.16 to 1.86$\pm$0.04 
as [Fe/H] varies from -1 to +0.5 dex. It demonstrates
that low-mass stars make up a larger fraction in metal-rich environments than in metal-poor ones. We performed simulations to assess the impact of unresolved binaries on the IMF power-law indices.  
After correction, the binary-adjusted $\alpha$ values retained a similar metallicity-dependent trend. 
Furthermore, by examining the IMF of the aggregate
sample,
we found the corrected indices ($\alpha_{\rm{1,corr}} = 1.48 \pm 0.03$ , $\alpha_{\rm{2,corr}} = 2.17 \pm 0.03$) are consistent with Kroupa's IMF values ($\alpha_1 = 1.3 \pm 0.5$ and $\alpha_2 = 2.3 \pm 0.3$). Finally, we verified the robustness of our results by testing different break points and mass bin sizes, confirming that the IMF’s dependence on [Fe/H] remains consistent.

\end{abstract}

\begin{keywords}
stars: main-sequence stars, metallicity, stellar initial mass function  -- methods: statistical, broken power-law

\end{keywords}



\section{Introduction}

The stellar initial mass function (IMF) describes the mass distribution of stars that formed simultaneously. Studies of the stellar IMF provide important constraints on the formation and evolution of stars, stellar populations, stellar clusters, and galaxies \citep[e.g.,][]{Cor-2005ASSL..327.....C, Li-2006AAS...20919803L, Kroupa-2008mru..conf..227K, Lee-2020SSRv..216...70L}. Therefore, the stellar IMF is essential for many fields of research in astrophysics.

\citet{Salpeter-1955} originally proposed that the stellar IMF can be approximated by a power-law distribution, as
\begin{equation}
    \xi (m)=\frac{\mathrm{d}n}{\mathrm{d}m} = Cm^{-\alpha}.
	\label{eq:IMf_sal}
\end{equation}
where $m$ and $n$ are the stellar mass and the corresponding stellar number, respectively. $C$ is the normalization constant. He derived an IMF power-law index of $\alpha$=2.35 ($\alpha= -\, \mathrm{d}\ln(\mathrm{d}n/\mathrm{d}m) / \mathrm{d}\ln (m) $). \citet{Scalo-1986FCPh...11....1S} was the first one to propose a comprehensive IMF with a broken power-law form, explicitly dividing the mass range into distinct segments with different slopes. Subsequently, a popular broken power-law IMF was developed by \citet{Kroupa-2001MNRAS.322..231K}. That is, the power-law indices are -0.7 $\pm$ 0.7, 1.3 $\pm$ 0.5, and 2.3 $\pm$ 0.3 for stars with mass ranging [0.01,0.08), [0.08,0.5), and $\geq$ 0.5 \M, respectively.

The IMF is often treated as universal; that is, stars are assumed to form with the same mass distribution in every environment. However, many recent works have challenged the invariant stellar IMF and reported that the universal IMF struggles to explain the observed data in a wide variety of environments \citep[e.g.,][]{Dabringhausen-2009MNRAS.394.1529D, Cescutti-2011A&A...525A.126C,Adams-2013hst..prop13232A,Bekki-2013ApJ...779....9B, Kalari-2018ApJ...857..132K, Yan-2024ApJ...969...95Y}. This controversy has motivated extensive investigations into potential correlations between IMF variations and environmental factors, like the star formation density, galactic velocity dispersion, and metallicity \citep[e.g.,][]{Dabringhausen-2011ASPC..440..261D, Kroupa-2013pss5.book..115K, Weidner-2013MNRAS.436.3309W, Lagattuta-2017ApJ...846..166L, Clauwens-2016MNRAS.462.2832C}. Notably, many studies have observationally confirmed that the IMF varies across different astrophysical environments \citep[e.g.][]{Cappellari-2012Natur.484..485C, Conroy-2012ApJ...760...71C, Li-2017ApJ...838...77L, Zhang-2018Natur.558..260Z, Zhou-2019MNRAS.485.5256Z, Hallakoun-2021MNRAS.507..398H, Yasui-2023ApJ...943..137Y, Yang-2024MNRAS.530.4970Y}.

The star formation rate (SFR) has been consistently demonstrated to contribute to the IMF variations \citep[e.g.,][]{Lee-2009ApJ...706..599L, Cappellari-2013MNRAS.432.1862C}. \citet{Weidner-2005ApJ...625..754W} focused on the integrated galactic initial mass function (IGIMF) of a galaxy. Their result shows a steeper slope of IGIMF than the IMF of \citet{Kroupa-2001MNRAS.322..231K}. They inferred a link between a galaxy’s SFR and the mass of its most massive young cluster, which consequently leads to a connection with the slope of the IGIMF. A top-heavy IMF in galaxies with high SFR was reported in \citet{Zhang-2018Natur.558..260Z}. \citet{Je-2018A&A...620A..39J} applied a galaxy-wide IMF model to study the possible reasons for the IMF variation; they demonstrated that the IMF variation correlates with metallicity, SFR, and age. 

Metallicity is also considered to be a factor related to the stellar IMF \citep{Villaume-2017ApJ...850L..14V, Meyer-2019ApJ...875..151M, Hallakoun-2021MNRAS.507..398H, Maksymowicz-2024MNRAS.531.2864M}.
\citet{Larson-2005MNRAS.359..211L} suggested that a top-heavy IMF should exist in the low-metallicity environment by analysing the far-infrared background radiation in starburst regions. 
\citet{Sharda-2023MNRAS.518.3985S} found that the IMF characteristic mass ($M_{c}$) is sensitive to changes in the [O/H] for the cloud model with low velocity dispersion and low pressure. 
Furthermore, \citet{Elmegreen-2008ApJ...681..365E} proposed that the characteristic mass depends weakly on the metallicity of young stars. The IMFs are more bottom-light than those of the Milky Way disc, as displayed in 
\citet{Gennaro-2018ApJ...855...20G}, who investigated the IMFs of stars with masses ranging from 0.45 to 0.8 \M\ of six ultra-faint dwarf Milky Way satellites. They found that the stellar IMF well correlates with the galaxy mean metallicity, i.e., a more bottom-light IMF for the galaxy with lower metallicity. The top-heavy IMF is expected in the clusters with low metallicity and larger pre-Galactic Center cloud density, as shown in \citet{Marks-2012MNRAS.422.2246M}. Collectively, these studies imply that metallicity modulates the IMF.

In addition, some other factors that contribute to the shape of the IMF are also explored, such as age, gas density, velocity dispersion and so on \citep[e.g.,][]{Cheng-2023MNRAS.526.4004C, Elmegreen-2004MNRAS.354..367E, Kobayashi-2010AIPC.1240..123K, van-2012ApJ...760...70V, Spiniello-2016ASSP...42..219S, Barber-2019MNRAS.483..985B}. The evidence for a dependence of the peak mass of IMFs on the local gas density was found by \citet{Levine-2006PhDT........29L}. By measuring the equivalent widths for luminous red galaxy spectra, \citet{Spiniello-2012ApJ...753L..32S} found that the low-mass end of IMF correlates with the age and metallicity. 
\citet{Geha-2013ApJ...771...29G} presented that the IMF power-law index of stars with a mass range of 0.5 to 0.8 \M\ becomes shallower with decreasing galactic velocity dispersion and metallicity. 
\citet{Dickson-2023MNRAS.522.5320D} studied 37 Milky Way globular clusters and presented that the IMFs of low-mass stars (<1 $M_{\odot}$) are strongly dependent on the dynamical age of clusters, whereas the high-mass IMF is not. \citet{Tanvir-2024MNRAS.527.7306T} concluded that the surface density is more likely responsible for the IMF variations compared to metallicity in early-dwarf galaxies.

Overall, many studies have explored the factors affecting the IMF using the data from star clusters or galaxies. However, only a limited number of field stars are available for studying the stellar IMF by directly counting them \citep[e.g][]{Scalo-1986FCPh...11....1S, Cignoni-2002ASPC..274..408C, Chabrier-2003ApJ...586L.133C, Best-2018PhDT.......159B}. 
 \citet{Hallakoun-2021MNRAS.507..398H} analyzed the stars within 250 pc selected from Gaia DR2 to investigate the stellar IMF; they found that the IMF of the blue halo ([M/H]<-0.6 dex) is bottom-heavy. However, for the thin-disc population, the IMF is similar to that of \citet{Kroupa-2001MNRAS.322..231K}. It suggests that the IMF depends on the environment in which the stars formed. Recently, \citet{Li-2023Natur.613..460L} (hereafter Li23) used $\sim$90,000 field M dwarf stars with masses ranging from 0.3 to 0.7 \M\ and distances spanning 100-300 pc to explore the stellar IMF as a function of metallicity. They developed a hierarchical Bayesian model for the vertical number density profile in the Milky Way based on a single power-law IMF. 
 Their results show that the power-law index of the IMF systematically increases with metallicity. However, Li23 used a narrow mass range and a relatively small sample size. 
In this work, we use over 500,000 dwarfs, spanning masses from 0.25 to 1.0 \M\ and distances of 150 to 350 pc, a sample significantly larger than that used by Li23, 
to further investigate the stellar IMF as a function of metallicity. 

This paper is structured as follows: Section \ref{sec:data} describes the determination of stellar atmospheric parameters and masses. We describe the method in Section \ref{sec:method}. It includes the selection function correction, the determination of stellar IMF, and the corresponding power-law index. The results and discussion are presented in Section \ref{sec:result_all} and \ref{sec:Discussion}, respectively. Finally, we draw conclusions in Section \ref{sec:Conclusion}.

\section{Data} \label{sec:data}

The data used in this work are taken from the ninth Data Release of LAMOST (LAMOST DR9 \footnote{https://www.lamost.org/dr9/v2.0/}). We focus on the stars with masses $\leq$ 1 \M, which are predominantly G, K, and M dwarfs. To obtain a complete sample of stars with masses $\leq$ 1 \M, we also include F-type stars from the LAMOST AFGK star catalogue. The M dwarfs are selected from the gM, dM, and sdM star catalogue. The determinations of atmospheric parameters and mass of each star are described in Section \ref{subsec:metallicity} and \ref{subsec:mass}, respectively. 

\subsection{Atmospheric parameters of dwarf stars} \label{subsec:metallicity} 

LAMOST DR9 provides precise metallicities for F, G, and K dwarfs but not for M dwarfs. Estimating the metallicity of M dwarfs is more challenging because their spectra, which are dominated by complex molecular bands, cannot be reproduced precisely by existing atmospheric models. Fortunately, the two components of a wide binary system are assumed to have the same metallicity. Therefore, it is feasible to calibrate the \feh\ of M dwarfs using F, G, or K dwarf companions \citep[e.g.,][]{Birky-2020ApJ...892...31B, Qiu-2024}. In \citet{Qiu-2024}, we identified 1308 LAMOST FGK+M wide binaries based on the catalogue of \citet{Badry-2021} to calibrate the \feh\ of M dwarfs. 

\begin{figure*}
\centering
\includegraphics[width=0.8\textwidth, trim=0cm 0.0cm 0.0cm 0.0cm, clip] {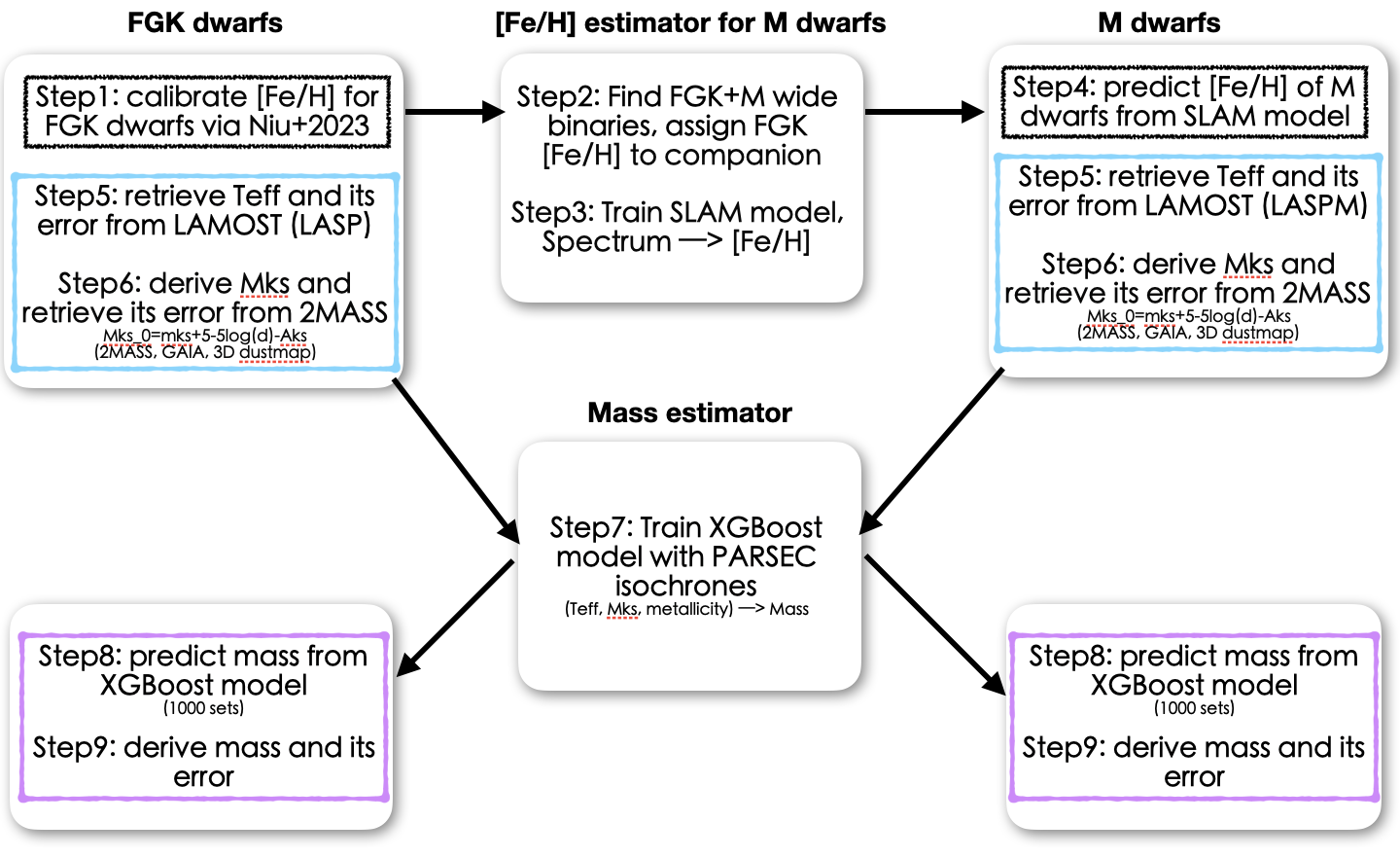}
\caption{The flowchart outlines the procedure for determining stellar masses. Detailed descriptions of each step are provided in subsections \ref{subsec:metallicity} and \ref{subsec:mass}.}\label{fig:process}
\end{figure*}

However, \citet{Niu-2023ApJ...950..104N} selected 2,296 FGK+FGK dwarf wide binaries from the LAMOST AFGK star catalogue. They investigated the \feh\ of both components in each binary system, where the \feh\ were derived from LAMOST Stellar Parameter pipeline \citep[LASP,][]{Wu-2011A&A...525A..71W}, and found that the estimations of \feh\ for A/F/G/K stars systematically depend on the effective temperature. To address this, they developed a broken power-law model to calibrate the \feh\ for stars with 4000 < \teff\ < 7000 K, as follows:

\begin{equation}
 \Delta \feh\ = \left\{
    \begin{aligned} 
    0.358 *(T_{\rm eff} /5281.4) ^{-2.404}-0.4 &,\, 5281.4 \leq T_{\rm eff}  \\
    0.358 *(T_{\rm eff} /5281.4) ^{1.254}-0.4 &,\,\, 5281.4> T_{\rm eff}\\
    \end{aligned}
 \right.
 \label{eq:deltafeh}
\end{equation}
where \teff\ is the effective temperature derived from the LASP model. 
Therefore, we first calibrated the \feh\ of F, G, or K dwarfs based on equation (\ref{eq:deltafeh}), as step 1 in Figure \ref{fig:process}. Then we used these calibrated \feh\ as reference values to calibrate the \feh\ of 1308 M dwarf secondaries (step 2). 

\begin{figure}
\centering
\includegraphics[width=0.5\textwidth, trim=0.0cm 0.0cm 0.0cm 0.0cm, clip] {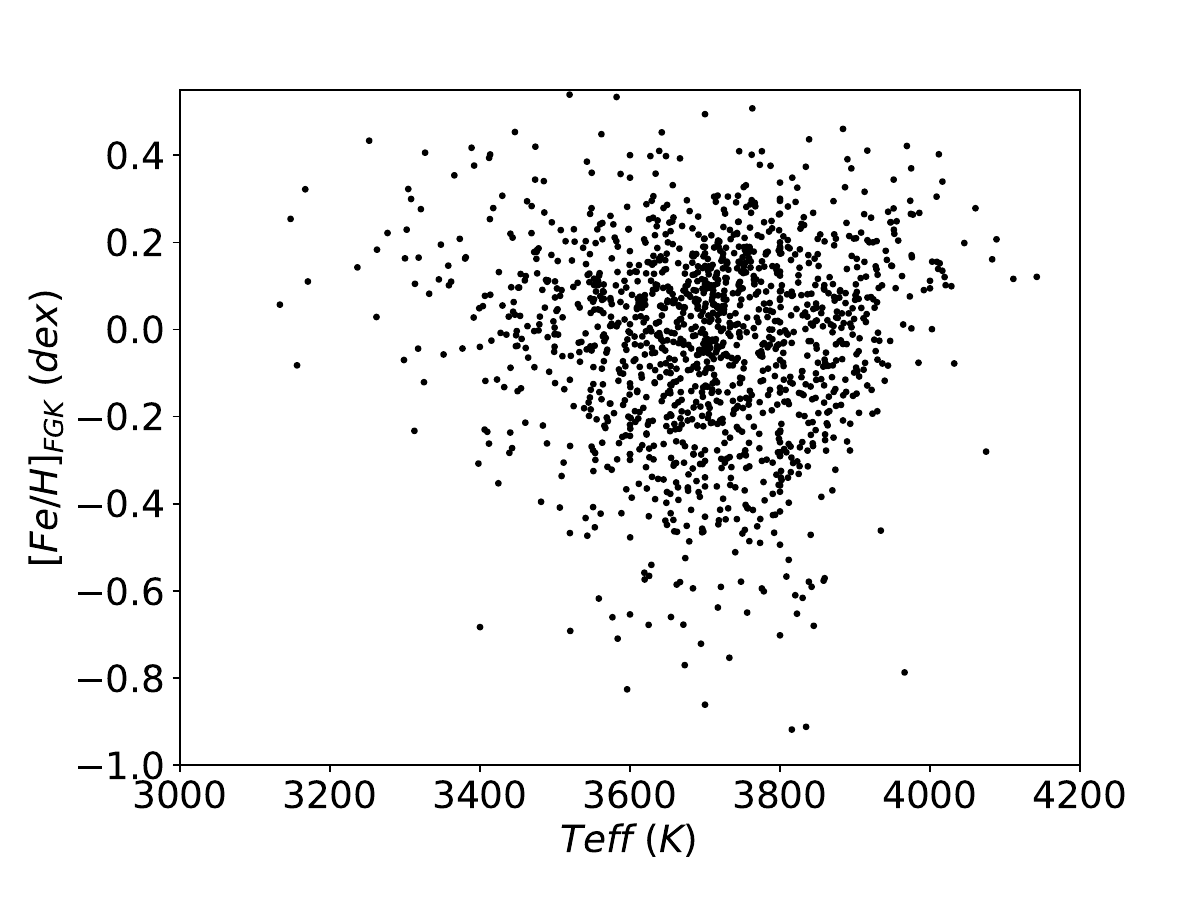}
\caption{The metallicity versus the effective temperature of 1308 LAMOST M dwarfs. The metallicities, calibrated with Equation (\ref{eq:deltafeh}), are inherited from the F, G, or K dwarf companions, whereas the \teff\ are taken from the LASPM pipeline.}\label{fig:teff_feh}
\end{figure}

As in our previous work \citep{Qiu-2024}, we trained a data-driven model, Stellar LAbel Machine \citep[SLAM,][]{Zhang-2020ApJS..246....9Z}, with 1000 LAMOST M dwarf low-resolution (R$\sim$1800) spectra and the corresponding calibrated \feh\ from F, G or K companions (step 3). The remaining 308 M dwarfs are regarded as the test set. The distribution of the calibrated \feh\ ($\rm [Fe/H]_{FGK}$) and the \teff\ of M dwarfs is shown in Figure \ref{fig:teff_feh}. The \teff\ is derived from the LAMOST stellar parameter pipeline of M-type stars \citep[LASPM,][]{Du-2021RAA....21..202D}. Apparently, most M dwarfs in the training set (95\%) have metallicities larger than -0.6 dex.  

\begin{figure*}
\centering
\includegraphics[width=1\textwidth, trim=0.0cm 0.0cm 0.0cm 0.0cm, clip] {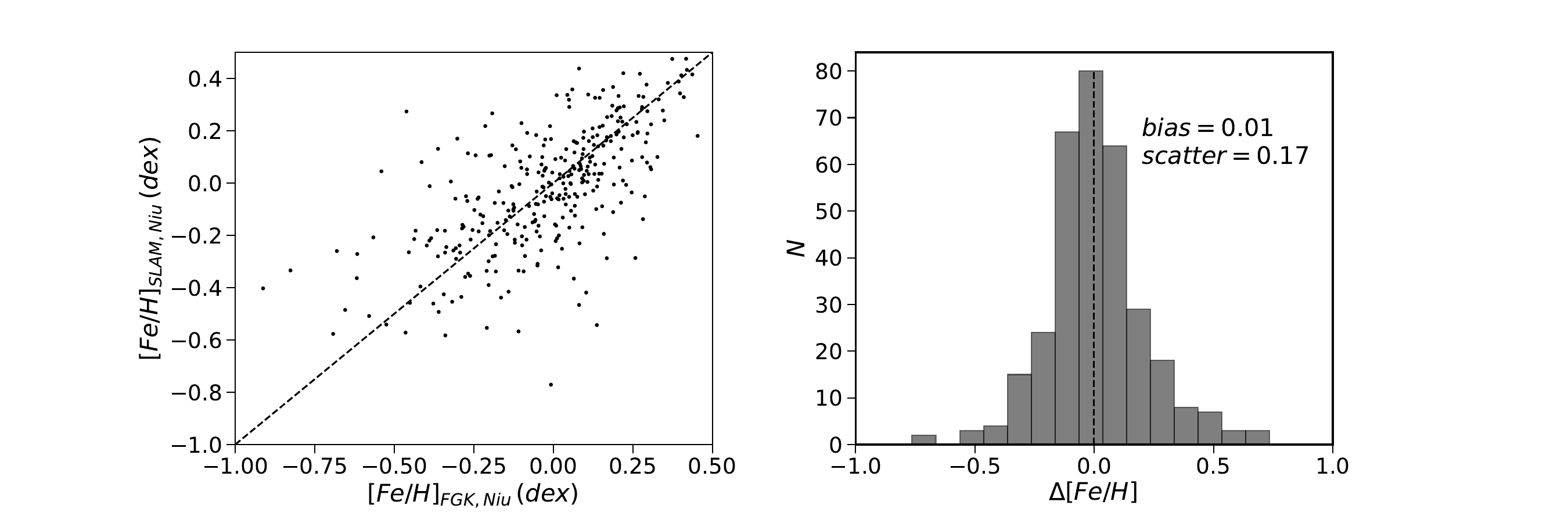}

\caption{The left panel shows the comparison in \feh\ of 308 test M dwarfs between the reference values $\rm [Fe/H]_{FGK}$ and the SLAM predictions $\rm [Fe/H]_{SLAM}$. The right panel presents the distribution of the differences, i.e., $\rm \Delta [Fe/H]$= $\rm [Fe/H]_{FGK}-[Fe/H]_{SLAM}$. Its mean and standard deviation values are 0.01 and 0.17, respectively.}\label{fig:test_feh}
\end{figure*}

In Figure \ref{fig:test_feh}, we compared the SLAM model predicted metallicity ($\rm [Fe/H]_{SLAM}$) with the reference values ($\rm [Fe/H]_{FGK}$) of the test set. The mean value of the bias is 0.01 with a scatter of 0.17 dex. We applied the SLAM model to all LAMOST M dwarf spectra to derive their \feh\ (step 4). The uncertainties of the predicted \feh\ can reach 0.15 dex for stars with signal-to-noise ratio in the $i$ band ($snri$) larger than 100. 

\begin{figure*}
\centering
\includegraphics[width=1\textwidth, trim=0.0cm 1.0cm 0.0cm 0.0cm, clip] {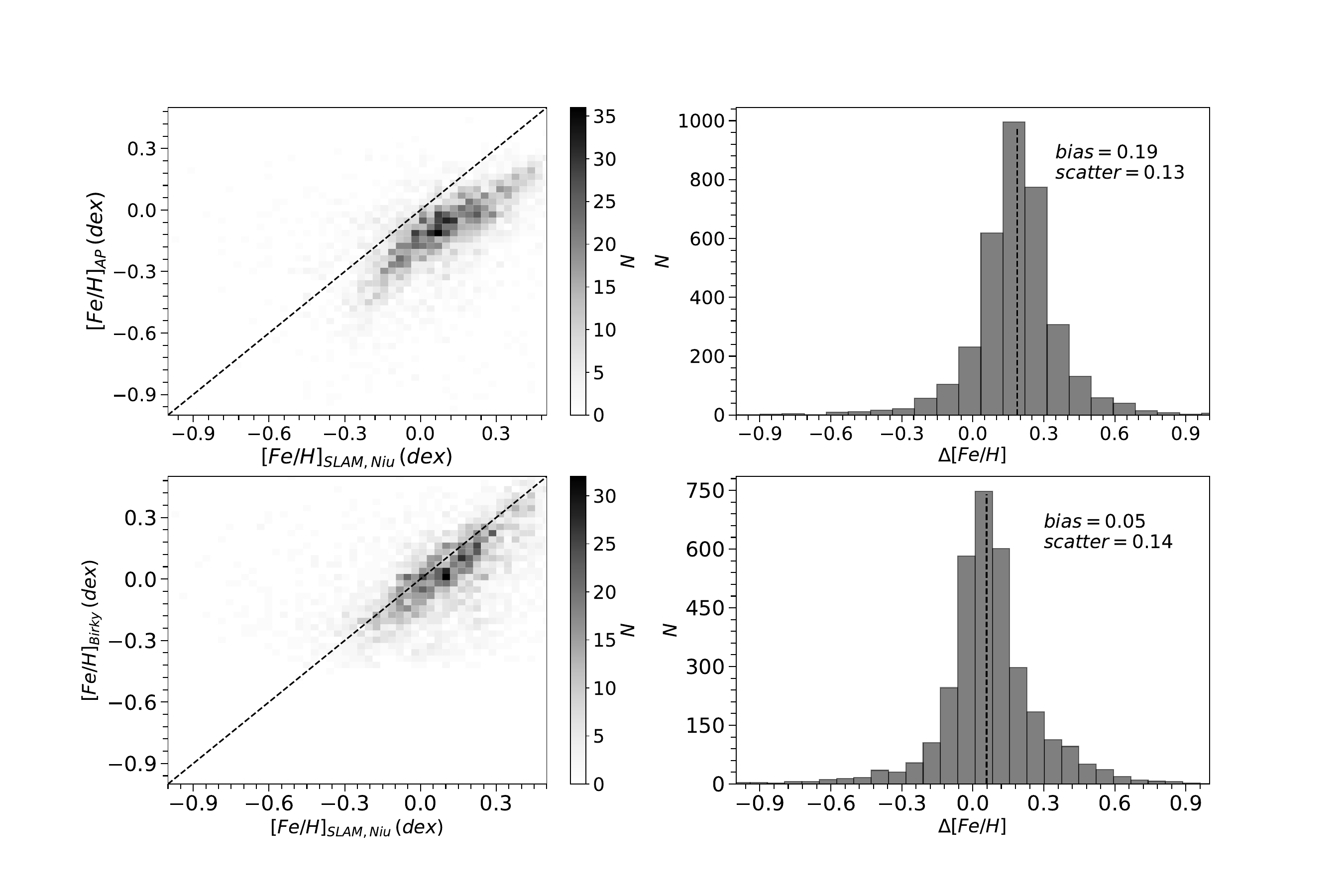}

\caption{The top-left panel shows the comparison between [Fe/H]$\rm_{SLAM}$ and those determined from APOGEE DR17 ([Fe/H]$\rm_{AP}$). The black dashed line is the one-to-one relation. The grayscale (white → dark gray) encodes the stellar number density in each [Fe/H]$\rm_{SLAM}$ and [Fe/H]$\rm_{AP}$ bin, with darker tones indicating higher densities. The corresponding histogram of the metallicity difference $\rm \Delta [Fe/H] (=[Fe/H]\rm_{SLAM}-[Fe/H]\rm_{AP})$ is displayed in the top-right panel. The bottom two panels are the same as the top two panels, but use the metallicities of \citet{Birky-2020ApJ...892...31B} as the reference. }\label{fig:delta_feh}
\end{figure*}

The validations of the SLAM \feh\ are shown in Figure \ref{fig:delta_feh}. There is a bias of 0.19 with a scatter of 0.13 dex in \feh\ compared with 3443 APOGEE DR17 M dwarfs \citep{Abdurro-2022ApJS..259...35A}. \citet{Souto-2022ApJ...927..123S} determined the chemical abundances of 11 M dwarfs by analysing the high-resolution near-infrared $H$-band spectra from the SDSS-IV/APOGEE survey \citep{Blanton-2017AJ....154...28B} and the synthetic spectra. They found a systematic offset of [Fe/H] = 0.24$\pm$0.11 dex compared with that of APOGEE DR16 \citep{J-2020AJ....160..120J}, which is similar to our results. Additionally, the bias is only 0.05 with a scatter of 0.16 dex between SLAM \feh\ and that of \citet{Birky-2020ApJ...892...31B}, who also calibrated the \feh\ of M dwarfs using their corresponding F, G, or K dwarf companions. It should be noted that the M dwarf stars with predicted \feh\,$<$ -0.6 dex may have larger uncertainties since the training samples falling into this metallicity range are fewer, as shown in Figure \ref{fig:teff_feh}. 


\subsection{Determination of stellar mass} \label{subsec:mass}

We used the effective temperature, metallicity, and absolute magnitude in 2MASS $K_s$ band to estimate the mass of each star. The \feh\ of F, G, and K dwarfs is calibrated by \citet{Niu-2023ApJ...950..104N} whereas that of M dwarfs is derived from the SLAM model. The \teff\ values for F, G, and K dwarfs are taken from LASP whereas those of M dwarfs are taken from LASPM (step 5). 

We obtained the $K_s$ band magnitudes of each star by cross-matching LAMOST F, G, K, and M dwarfs with 2MASS \citep{Skrutskie-2006AJ....131.1163S}. We retrieved the reddening value ($E(B-V)$) of each star from the three-dimensional dust map \citep{Green-2019ApJ...887...93G}. Adopting $ A_V=3.1\cdot E(B-V)$ and $ A_{K_s}=0.078\cdot A_V$ \citep{Wang-2019ApJ...877..116W}, the extinction-corrected absolute magnitude in the $K_{s}$ band is $M_{K_s0}=M_{K_s}-A_{K_s}$, where $M_{K_s}=K_s+5-5\cdot log_{10}(D)$, D is the distance of the star in pc, adopted from \citet{Bailer-2021AJ....161..147B} (step 6). 

We trained an XGBoost model \citep{Chen-2016arXiv160302754C}, a tree-based machine learning algorithm, with a training dataset (metallicity, temperature, $M_{K_s}$, mass) that comes from the PARSEC isochrones \citep{Bressan-2012MNRAS.427..127B, Chen-2014MNRAS.444.2525C}, like step 7 in Figure \ref{fig:process}. We then used the trained model to derive the mass of all dwarfs with known \feh, \teff\, and $M_{K_s0}$. The uncertainties of \feh\ and \teff\ for F, G, and K dwarfs are taken from the LASP. For M dwarfs, \feh\ uncertainties are derived from the SLAM model, while \teff\ uncertainties come from the LASPM. 
We propagate 2MASS $K_s$ photometric errors into the $M_{K_s0}$ uncertainty for all stars, which is reasonable given the precise distances and extinctions in the solar neighbourhood.
For each star, we randomly sample 1000 sets of parameters (\feh, \teff, and $M_{K_s0}$) from their distributions and derive the mass from the XGBoost model for each set independently (step 8). The mean and standard deviation values of 1000 predicted masses are adopted as the stellar mass and the corresponding uncertainties of the star (step 9). 

\begin{figure*}
\centering
\includegraphics[width=1\textwidth, trim=0.0cm 1.0cm 0.0cm 0.0cm, clip] {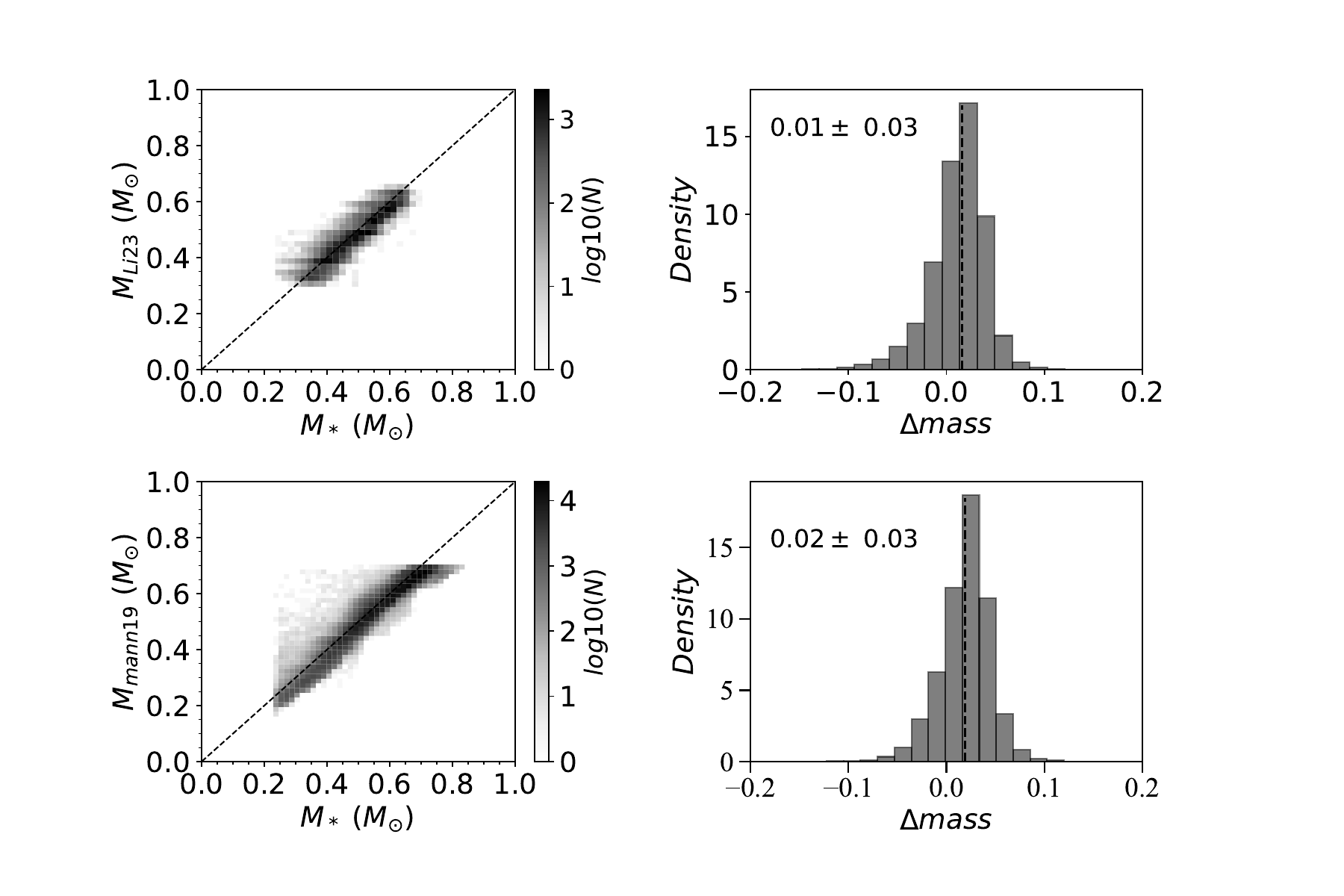}

\caption{The top-left panel shows the mass comparison between our work ($M_*$) and that of Li23 ($M_{Li23}$). A white-to-dark-gray scale encodes the logarithmic stellar counts within each $M_*$–$M_{Li23}$ bin. The distribution of mass difference $\Delta \rm{mass}=M_*-M_{Li23}$ is displayed in the top-right panel. The two bottom panels are the same as the top ones, but for the mass comparison between our work and that of \citet{Mann-2019ApJ...871...63M} ($M_{\rm{mann19}}$). }\label{fig:mass_com}
\end{figure*}

The top two panels in Figure \ref{fig:mass_com} show the comparison of masses between our work and those of Li23. It shows that our masses agree with those of Li23, exhibiting a bias of 0.01$\pm$0.03 \M. This is expected since Li23 used the same method to derive the stellar mass. \citet{Mann-2019ApJ...871...63M} used 62 nearby binaries to establish an empirical relationship between luminosity ($M_{K_s}$) and stellar mass. This relationship is applicable to stars with the mass spanning 0.075 < $M_*/M_{\odot}$ < 0.70. We compared our masses with those of \citet{Mann-2019ApJ...871...63M}, as shown in the two bottom panels of Figure \ref{fig:mass_com}. The result exhibits an offset of 0.02 with a scatter of 0.03 \M. It indicates that the masses in this work are in good agreement with those of \citet{Mann-2019ApJ...871...63M}.

\subsection{Volume completeness} \label{subsec:volumn}

\begin{figure}
\centering
\includegraphics[width=0.5\textwidth, trim=0.0cm 0.0cm 0.0cm 0.0cm, clip] {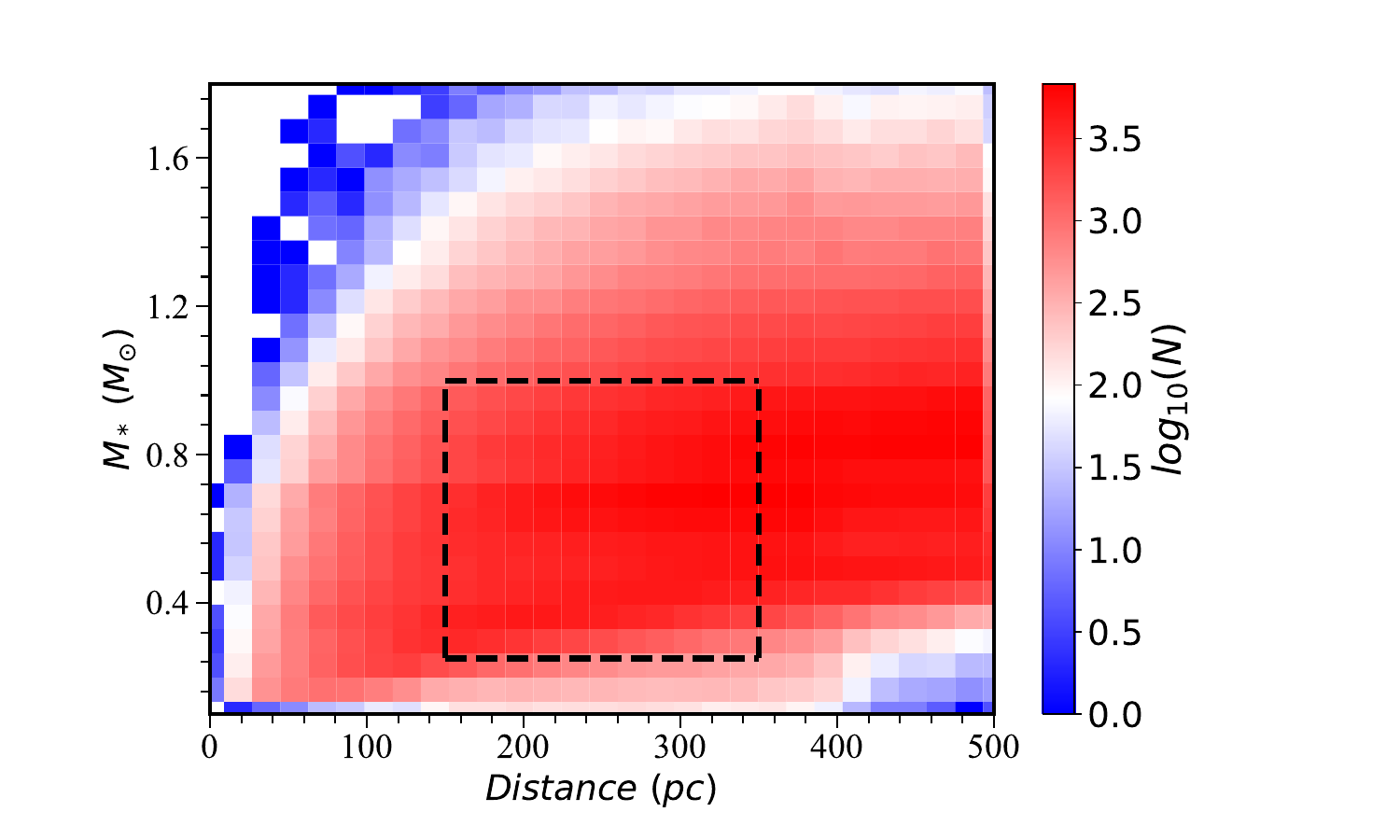}
\caption{The stellar mass versus distance of LAMOST dwarfs. The colours encode the logarithmic values of stellar density in each mass and distance bin. The subsample analysed in this work with 150 $<$ distance $<$ 350 pc and $\rm 0.25 \leq M_*/M_{\odot} \leq 1$ are highlighted in the black box.}\label{fig:mass_distance}
\end{figure}

 We focus on the IMF of stars with masses $\leq$1 \M. 
 Most of these stars have survived for nearly the entire age of the Universe. Meanwhile, their masses have changed negligibly since birth—whether through stellar winds or binary interactions.
  The masses derived in Section \ref{subsec:mass} can therefore be regarded as initial masses. To minimise the Malmquist effect \footnote{The Malmquist effect is a bias in the measurement of astronomical objects, particularly related to the way in which brighter objects are more likely to be included in observational samples as distance increases. This effect arises because only the brighter objects at a greater distance can be observed due to the limiting sensitivity of the observational instruments.}, we limit our sample to 0.25 $\leq$ mass $\leq$ 1.0 $M_{\odot}$ and 150 < distance < 350 pc (black box in Figure \ref{fig:mass_distance}). This selection yields a data set of more than 500,000 dwarf stars.

\section{Method}\label{sec:method}
Accounting for survey incompleteness is one of the main challenges in deriving the stellar IMF. We correct the observed field star number densities with the selection-function formalism of \citet{Liu-2017RAA....17...96L}, as outlined in Section \ref{sec:select_function}. The procedure used to estimate the IMFs of stars with different \feh\ is described in Section \ref{sec:IMF}. In Section \ref{sec:power_law}, we model the IMFs with a broken power-law function. 

\subsection{Select Function Correction} \label{sec:select_function}

The LAMOST provides a limited number of stars due to its targeting strategy. A statistical method developed by \citet{Liu-2017RAA....17...96L}, which can be used to recover the selection function of the spectroscopic survey and derive the stellar number density of the Milky Way based on photometric colours and magnitude.  

The completeness of 2MASS in $K_{s}$ band \citep{Skrutskie-2006AJ....131.1163S} is 99\% for stars with LAMOST observed luminosity limitation. Assuming that, for a given set of Galactic coordinates ($l$, $b$) and distance D, the selection of the LAMOST spectroscopic targets is determined solely by the colour–magnitude diagram. 
Therefore, the photometric (ground truth) stellar number density profile ($\nu_{\rm ph}$) can be recovered from that of spectroscopic data by correcting the selection function, i.e.,
\begin{equation}
\nu_{\rm ph}(D|l,b,c,m)=\nu_{\rm sp}(D|l,b,c,m)\cdot S^{-1}(l,b,c,m).
\label{eq:sp_ph}
\end{equation}
where $c$ and $m$ are the colour and magnitude of stars, respectively. That is, $c=J-K_{s}$ and $m=K_{s}$, which come from 2MASS in this work. $S$ can be determined as

\begin{equation}
\begin{split}
S(l,b,c,m)&=\frac{\int_0^{\infty} \nu_{\rm sp}(D|l,b,c,m) \Omega D^2 dD}{\int_0^{\infty} \nu_{\rm ph}(D|l,b,c,m) \Omega D^2 dD}\\
&=\frac{n_{\rm sp} (l,b,c,m)}{n_{\rm ph} (l,b,c,m) }.
\label{eq:s}
\end{split}
\end{equation}
where $\Omega$ is the solid angle associated with the line-of-sight. $\nu_{\rm ph}$ and $\nu_{\rm sp}$ are the photometric and spectroscopic stellar density distributions of a given ($l, b, c, m$), respectively. $n_{\rm ph}$ and $n_{\rm sp}$ are the numbers of photometric and spectroscopic stars with specific $l, b, c$ and $m$, respectively.

The photometric stellar density distribution of a given ($l,b$) can be obtained by integrating over colour index and magnitude,
\begin{equation}
\nu_{\rm ph}(D|l,b)=\iint \nu_{\rm sp}(D|l,b,c,m)S^{-1}(l,b,c,m)dcdm.
\label{eq:all_s}
\end{equation}
And a kernel density estimation can be used to derive the $\nu_{\rm{sp}}$.

Similar to Equation (\ref{eq:all_s}), for subsample $F$, selected from the spectroscopic data under specific selection criteria, the corresponding stellar profile of photometric data is
\begin{equation}
\nu_{\rm ph}(D|F,l,b)=\iint \nu_{\rm sp}(D|F,l,b,c,m)S^{-1}(l,b,c,m)dcdm.
\label{eq:sub}
\end{equation}
The details about the correction of the selection function refer to Section 2 in \citet{Liu-2017RAA....17...96L}. 

\begin{figure*}
\centering
\includegraphics[width=1.\textwidth, trim=0.0cm 0.0cm 0.0cm 0.0cm, clip] {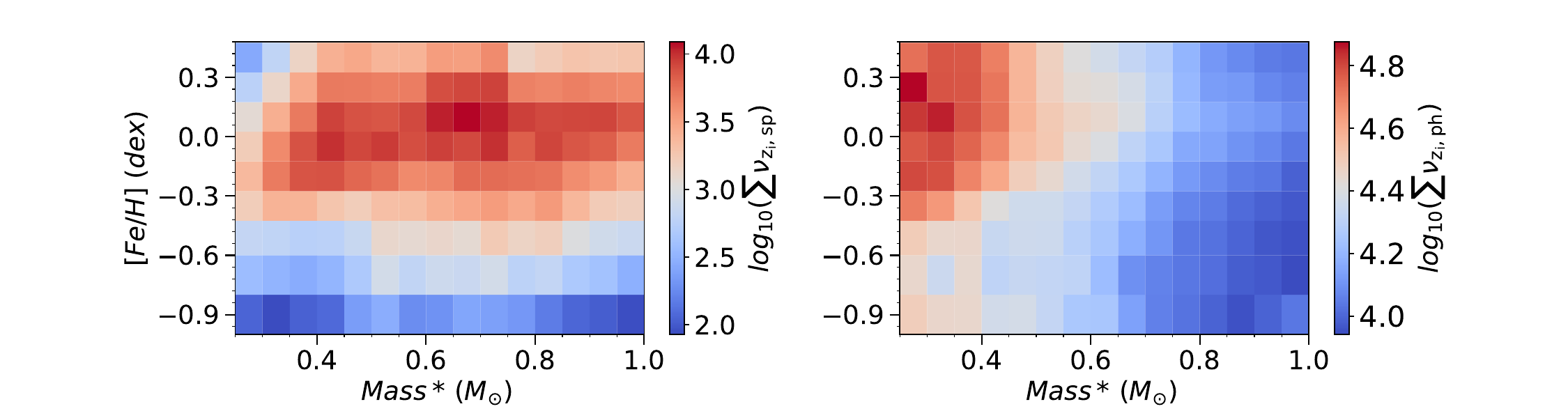}
\caption{The left panel represents the metallicity-mass diagram, where colours encode the total observed number density of LAMOST stars in each [Fe/H] and mass bin. $\nu_{\rm z_{i},sp}$ is the median $\nu_{\rm{sp}}$ of stars with a given metallicity and mass at the $i$-th vertical distance. The right panel is the same as the left panel but with colours representing the total corrected number density of stars. The estimations of $\sum \nu_{\rm z_{i},sp}$ and $\sum \nu_{\rm z_{i},ph}$ are described in Sections \ref{sec:select_function} and \ref{sec:IMF}, respectively. 
}\label{fig:obs_cor}
\end{figure*}

Figure \ref{fig:obs_cor} shows the \feh\ versus stellar mass. In the left panel, the colours represent the $\sum \nu_{\rm z_{i},sp}$, obtained by replacing $\nu _{\rm z_{i},sp}$ with $\nu _{\rm z_{i},ph}$ in Equation (\ref{eq:N_IMF}). In the right panel, the $\sum \nu _{\rm z_{i},ph}$ is directly derived from Equation (\ref{eq:N_IMF}).
The difference between two panels indicates that the selection function of LAMOST is more pronounced for stars with lower mass or lower metallicity than that at the higher mass and metal-rich end.

\subsection{Determination of stellar IMF} \label{sec:IMF}

We used a primitive binning method to explore the stellar IMFs across different \feh. First, we split the whole sample into subsamples based on stellar mass and \feh. Each subsample was further divided into vertical distance ($z$) bins. To reduce the Poisson noise and ensure a sufficient number of stars within each subsample, we set the mass bin width to 0.05 \M, defined metallicity bins as [-1, -0.8, -0.6, -0.45, -0.3, -0.15, 0, 0.15, 0.3, 0.5] dex, and set a vertical distance bin size of 40 pc, as $z\_bin$=[0, 40, 80, 120, 160, 200, 240, 280, 320] pc.

\begin{figure*}
\centering
\includegraphics[width=1.1\textwidth, trim=0.0cm 0.0cm 0.0cm 0.0cm, clip] {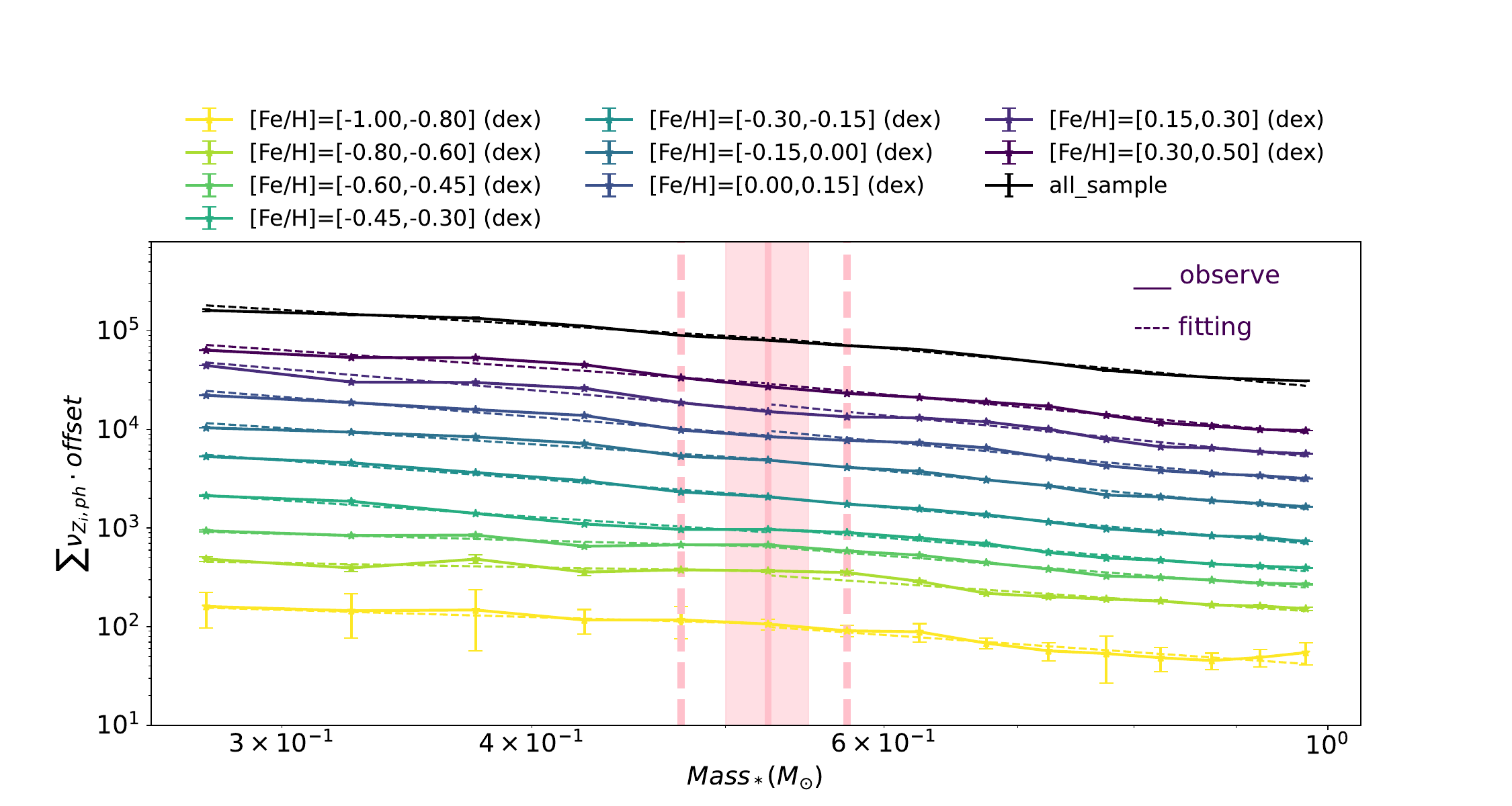}
\caption{Both axes are shown on logarithmic scales. The solid lines, transitioning from yellow to dark purple, represent the initial mass function of stars with metallicities ranging from –1.0 to +0.5 dex, the black solid line denotes the IMF derived from the full, unbinned sample. All the solid lines are computed with a mass bin size of 0.05\M. The dotted lines show the corresponding IMFs calculated with a coarser mass bin size of 0.1 \M. A vertical pink solid line marks the location of 0.525 \M, corresponding to the mass bin [0.50,0.55) \M\ (highlighted in pink). Two vertical pink dotted lines at 0.475 \M\, and 0.575\M\, indicate the adjacent bins, [0.45,0.50) and [0.55,0.60)\M, respectively.}\label{fig:dndm}
\end{figure*}

The $\nu_{\rm ph}$ of each star in a subsample with a given mass, metallicity, and $z$ was calculated as described in Section \ref{sec:select_function}.
We assumed that the stars within the thin disc follow a flat radial stellar density.
The total number of stars counted from a complete sample in each mass and \feh\, bin can be determined by


\begin{equation}
\begin{split}
N_{\rm ph}(m,\rm{[Fe/H]}) &\propto  \int_{0}^{\infty}\nu_{\rm ph}(m,\rm{[Fe/H]},z) dz\\
&\approx \sum_{i=1}^{8} \nu_{\rm z_{i},ph}(m,\rm{[Fe/H]},z_{i})\Delta z_{i},
\label{eq:N_IMF}
\end{split}
\end{equation}
where $\nu _{\rm z_{i},ph}$ denotes the median $\nu_{\rm ph}$ of stars in the $i$-th $z$ bin. Therefore, for a given [Fe/H], the distribution of the summed $\nu_{\rm z_{i},ph}$ values along $z$ in different mass bins represents the stellar IMF. The IMFs of stars with metallicities ranging from -1.0 to 0.5 dex are shown in Figure \ref{fig:dndm}, 
plotted as solid lines that transition in colour from yellow (metal-poor) to dark purple (metal-rich).

\subsection{The power-law model of stellar IMF}\label{sec:power_law}

We model the stellar IMFs with a broken power-law function. Because small-scale fluctuations obscure the break point—particularly in the metal-rich bins—and because metallicities below -0.6 dex carry larger uncertainties (Section \ref{subsec:metallicity}), we use the IMF for stars with [Fe/H]=[-0.6, -0.45) dex as a reference to locate the break point.

We adopted a multiple-order differences approach to estimate the local derivative of the stellar number distribution $f(m)$, as
\begin{equation}
f'(m_j) = \frac{-f(m_{j+2}) + 8f(m_{j+1}) - 8f(m_{j-1}) + f(m_{j-2})}{12 \Delta m}.
\label{eq:fx'}
\end{equation}
where $m_{j}$ is the centre of the $j$-th mass bin, specifically, 0.275, 0.325, ..., 0.975. $\Delta m$ is the mass bin size of 0.05 \M. 
The sign and magnitude of $f'(m_j)$ indicate the population’s sensitivity to mass variations, and sudden changes in $f'(m)$ may signal transitions such as break points in the initial mass function.

We calculated the multiple-order differences of that stellar IMF and obtained
the peak of $f'(m_j)$ at 0.525 \M. It marks the location of the break point, which is consistent with the commonly adopted value of 0.5 \M\ reported by \citet{Kroupa-2001MNRAS.322..231K}. 
While there is currently no physical explanation for this break point. We note that the break point identified in our analysis lies close to the transition between M dwarfs and F/G/K dwarfs. As our determination of [Fe/H] employs different methodologies for these two stellar populations, we cannot rule out the possibility that part of the observed break is induced by methodological differences.

We fit each IMF in Figure \ref{fig:dndm} with a broken power-law function using a simple Bayesian framework. Following the reference IMF, we adopt a break point of 0.525 \M\ for all IMFs. 

We assumed that the stellar IMF is
\begin{equation}
 \xi (m)= \left\{
    \begin{aligned} 
    C_1\cdot m^{-\alpha_1},& \,\, 0.25\leq m/M_{\odot} \leq 0.525 .\\
    C_2\cdot m^{-\alpha_2},&\, \, 0.525 < m/M_{\odot}\leq 1 .\\
    \end{aligned}
 \right.
 \label{eq:exp}
\end{equation} 
where $\alpha_1$ and $\alpha_2$ are the power-law indices for stars with mass $\leq$ 0.525 $M_{\odot}$ and $>$ 0.525 $M_{\odot}$, respectively. $C_1$ and $C_2$ are the corresponding normalization constants. 

For stars with mass $\leq$0.525 \M\ and a given \feh, the joint posterior distribution of the IMF parameters is
\begin{equation}
\begin{split}
p(\alpha_{1},C_{1}|\{m_j\},\{&N_{{\rm ph},j}\},\rm{[Fe/H]}) \propto \\&p(\alpha_{1},C_{1})\mathcal{L}\left(\{N_{{\rm ph},j}\}|\{ m_j\},\alpha_{1},C_1,\rm{[Fe/H]}\right).
\label{eq:IMF_joint}
\end{split}
\end{equation}
where $N_{{\rm ph},j}$ is the total photometric star count in the $j$-th mass bin (see Equation (\ref{eq:N_IMF})), with $j$=0, 1, …,5 corresponding to the mass bins of 0.25, 0.3, ..., 0.55\,\M.  We adopted uniform priors for $C_1$ and $\alpha_1$, with $C_1$ ranging from 5000 to 20000 and $\alpha_1$ varying from 0 to 4, respectively. The likelihood for the power-law model can be written as
\begin{equation}
\begin{split}
\mathcal{L}\left(\{N_{{\rm ph},j}\}|\{m_j\},\alpha_1,C_1,\rm{[Fe/H]}\right)= \prod \limits_{j=0}^{5} \exp \left(-\frac {C_1 m_{j}^{-\alpha_1}- N_{{\rm ph},j}}{2\sigma^2_{N_{{\rm ph},j}}}\right)^2,
\label{eq:IMF}
\end{split}
\end{equation}
where $\sigma_{N_{{\rm ph},j}}$ is the uncertainty of ${N_{{\rm ph},j}}$.

We derive $\alpha_1$ and $C_1$ for each IMF with a Markov-Chain Monte-Carlo (MCMC) sampler. $\alpha_2$ and $C_2$ are obtained in the same way, but ${N_{{\rm ph},j}}$ and $m_j$ are taken from stars with $m_j > 0.525$ \M\ (i.e., the mass bins of (0.55, 0.6, ... , 1) \M). 

%

\section{Results}\label{sec:result_all}
We investigate the IMF power-law indices for the full, unbinned sample and for stars in individual [Fe/H] in Subsection \ref{sec:power_feh}. The comparison of stellar IMF indices between our work and Li23 is presented in Subsection \ref{sec:comparison_Li23}.
\subsection{IMF Power-law indices as a function of \feh}\label{sec:power_feh}

First, we analyzed the IMF of the entire sample without dividing it into metallicity bins (black solid line in Figure~\ref{fig:dndm}). For the full sample, we derived slopes of $\alpha_1=1.19 \pm 0.03$ and $\alpha_2=1.81 \pm 0.03$. The dotted lines in figure ~\ref{fig:dndm} illustrate the corresponding two-segment fits.



\begin{figure}
\centering
\includegraphics[width=0.5\textwidth, trim=0.0cm 1.0cm 0.0cm 0.0cm, clip] {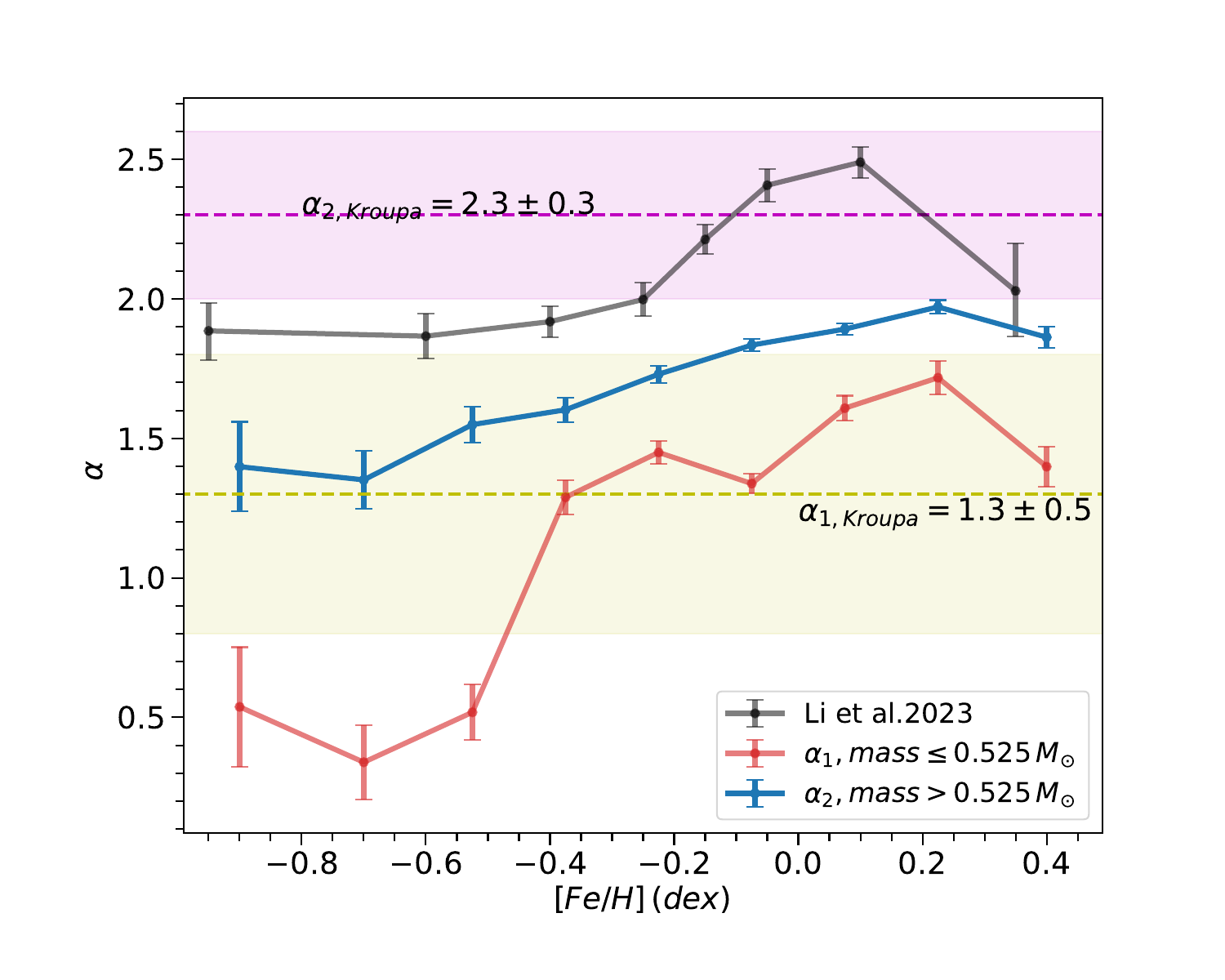}
\caption{The red and blue lines represent the IMF power-law indices as a function of \feh\ for stars with mass $\leq$ 0.525 \M ($\alpha_1$) and > 0.525 \M\ ($\alpha_2$), respectively. The black line displays the result of Li23 as a reference. The yellow dashed line marks the Kroupa's slope for stars with a mass range of [0.08, 0.5) $M_{\odot}$ ($\rm \alpha_{1,Kroupa}=1.3$), with the surrounding yellow shading indicating its quoted uncertainty ($\pm$0.5).
Likewise, the purple line and shading show the slop value and uncertainty for mass $\geq$ 0.5 \M\ ($\rm \alpha_{2,Kroupa} =2.3\pm 0.3$).  
 }\label{fig:alpha_feh}
\end{figure}

We then derived the power-law slopes for the IMF in each metallicity bin. Figure \ref{fig:alpha_feh} displays $\alpha_1$ (red) and $\alpha_{2}$ (blue) as a function of metallicity. The results show that both indices increase with metallicity, implying that over the entire mass range of [0.25, 1] \M, metal-rich stellar populations tend to produce a larger fraction of low-mass stars than metal-poor ones. Specifically, $\alpha_1$ varies from 0.54$\pm$0.21 to 1.40$\pm$0.07 as the metallicity changes from -1 to +0.5 dex. And the $\alpha_2$ changes from 1.40$\pm$0.16 to 1.86$\pm$0.04.

It is worth highlighting that there is a significant difference between $\alpha_1$ and $\alpha_2$ of metal-poor stars, particularly for stars with \feh\ $<$ -0.45 dex. In contrast, for metal-rich stars, the $\alpha_1$ is closer to $\alpha_2$, which explains the absence of a distinct break point in the IMFs for stars with high metallicity. There is as yet no known 
reason for this abrupt change. It may indicate that the break point itself shifts with \feh\ or the stellar IMFs of different metallicities follow distinct forms.

\subsection{Comparison with Li23}\label{sec:comparison_Li23}

Li23 developed a hierarchical Bayesian model based on the stellar photometric number density ($\rm \nu _{ph}$) to explore the IMF of stars with masses spanning 0.3 to 0.7 \M\ and distances in the range of 100-300 pc.
We compared our results with those of Li23 (black line), as shown in Figure \ref{fig:alpha_feh}. It shows that the trend in the variation of the power-law index in this work is similar to that of Li23.  
It is noteworthy that Li23 modeled the vertical distribution of stars in the Galactic disc with an exponential density profile, adopting a uniform scale height irrespective of stellar mass and metallicity. However, it was reported that the scale height ranges from 280–300 pc for early-type dwarfs and increases to about 350 pc for late-type dwarfs \citep{Siegel-2002ApJ...578..151S}. It indicates that Li23 may apply an oversimplified assumption in the model of the stellar density profile. 

In our analysis, we sum the star counts within each $z$ bin directly, without assuming an exponential profile, to derive the IMF. This difference in methodology likely explains the systematic offset between our power-law indices and those of Li23. Meanwhile, our mass range also differs from that of Li23, which may further contribute to the discrepancy. Moreover, unlike this study, Li23 did not introduce a break point in their analysis. The use of such a feature in our methodology may itself lead to differences in the results.



It is also noted that the $\alpha$ values of Li23 drop from 2.50 $\pm$ 0.06 to 2.00 $\pm$ 0.17 for stars with metallicity bin changes from [0, 0.2] dex to [0.2, 0.5] dex. Similarly, in our work, $\alpha_1$ changes from 1.72$\pm$0.06 to 1.40$\pm$0.07 as the metallicity changes from [0.15, 0.3) dex to [0.3, 0.5] dex, while $\alpha_2$ declines from 1.97$\pm$0.02 to 1.86$\pm$0.04. This variation may be attributed to the migration of stars from regions near the Galactic centre \citep[][Li23]{2015MNRAS.447.3526K}, which is composed of a complex population. 

\section{Discussion} \label{sec:Discussion}
Unresolved binaries would affect the stellar luminosity and thus the IMF, especially for low-mass stars \citep{Kroupa-2018arXiv180610605K}. We examined this effect with simulated data in Subsection \ref{sec:binaries}. The choice of break point and of the mass-bin width can also influence the derivation of IMF index \citep{Ma-2005ApJ...629..873M,Cara-2008ApJ...686..148C}; we set different break points and mass bin sizes to explore the robustness of the IMFs in Subsections \ref{sec:beark_point} and \ref{sec:mass_bin}, respectively. 

\subsection{Binaries correction}\label{sec:binaries}
\subsubsection{Simulation} \label{sec:simu_binaries}

Unresolved binaries must be taken into account when deriving the stellar IMF. If such a system is treated as a single star, the additional light from the secondary leads to an overestimate of the mass and, in turn, to a spurious bottom-light IMF. To quantify this bias, we construct a mock data set of 300,000 single stars drawn from a broken power-law mass function with indices $\alpha$=1.3 for stars with mass $\leq$ 0.525 \M\ and $\alpha$=2.3 for stars with mass > 0.525 \M, spanning a mass range of 0.1–2.0 \M. 

The binary fraction is defined as
\begin{equation}
f_b=\frac{N_{\rm{bin}}}{N_{\rm{bin}}+N_{\rm{sin}}}.
\label{eq:all}
\end{equation}
where $N_{\rm{bin}}$ and $N_{\rm{sin}}$ are the numbers of unresolved binaries and single stars, respectively.

We construct binary populations by randomly pairing stars in the simulated catalogue, without any dependence on primary mass or metallicity, to explore global binary fractions from 0\% to 60\%.  
This study does not account for binary evolution or for non-random mass-ratio distributions, factors that could in principle affect the derived IMF slopes.
Using the mass-to-luminosity ratio (LMR) of the PARSEC model, we convert the combined luminosity of each unresolved binary to its mass. We then use the same model described in Section \ref{sec:power_law} to derive the IMF power-law index of the sample that includes unresolved binaries. 

\begin{figure}
\centering
\includegraphics[width=0.5\textwidth, trim=0.0cm 0.0cm 0.0cm 0.0cm, clip] {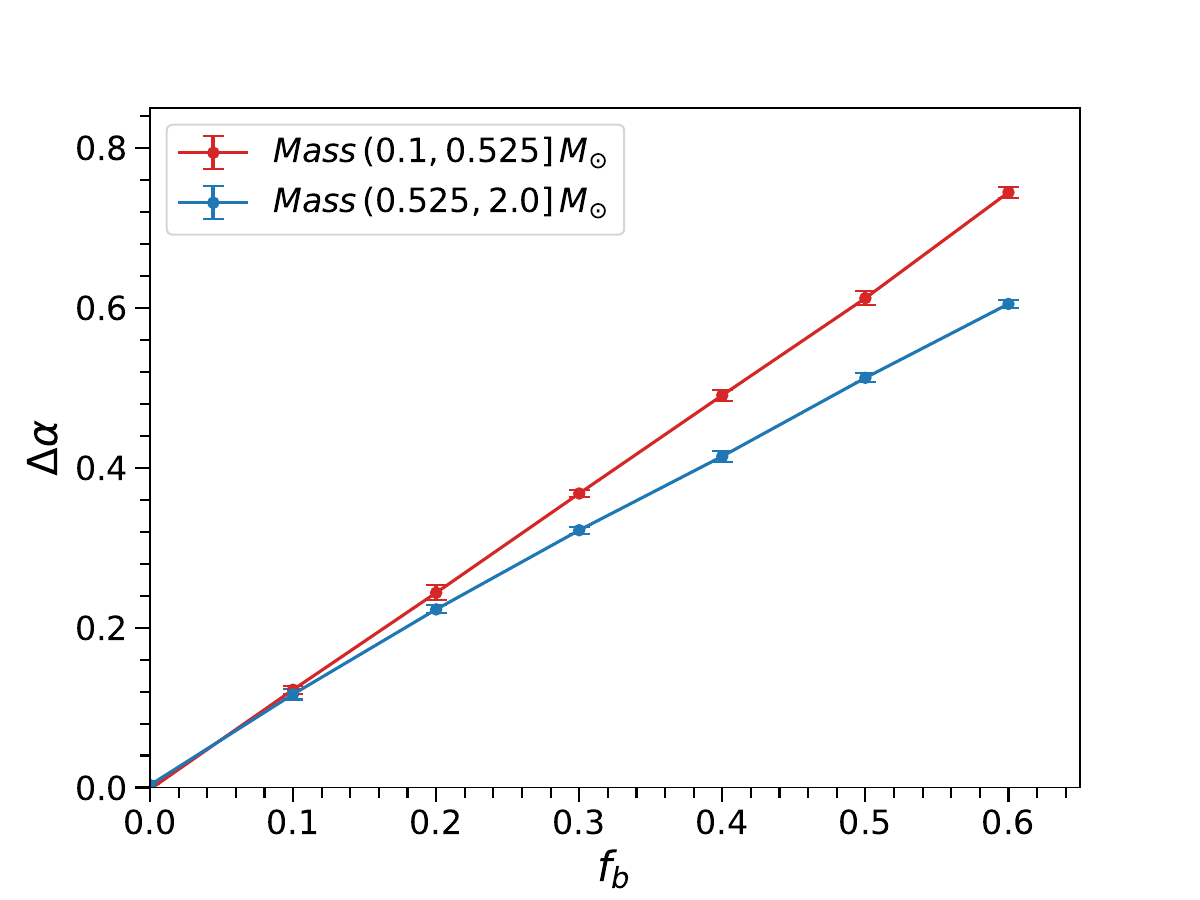}
\caption{The shift values of the power-law index ($\Delta \alpha$) as a function of binary fractions for stars with mass $\leq$ 0.525 \M\ (red) and > 0.525 \M\ (blue), respectively. As the binary fraction rises from 0\% to 60 \%, $\Delta \alpha$ grows from 0 to 0.74 for the low-mass subsample, but only from 0 to 0.60 for the high-mass subsample.Results are based on random pairing from a given mass function, independent of primary mass and metallicity.}\label{fig:fb}
\end{figure}

The deviations from the expected values of 1.3 (mass$\leq$0.525 \M) or 2.3 (mass>0.525 \M) represent the shift values due to unresolved binaries. Figure \ref{fig:fb} displays the shift values as a function of binary fractions. It indicates that reinstating the numerous low-mass companions that are missed in system counts increases the fitted index $\alpha$. Moreover, because random pairing produces a mass-ratio distribution that is heavily skewed toward small values \citep{Kouwenhoven-2009A&A...493..979K,Wang-2025arXiv250612987W}, most hidden companions fall below 
0.525 \M. As a result, the low-mass bins receive the largest fractional boost, steepening the fitted $\alpha$ most strongly at the low-mass end of the IMF \citep{Weidner-2009MNRAS.393..663W, Kroupa-2018arXiv180610605K}. That is, under the same unresolved binary fraction, the shift values are more pronounced for lower-mass stars.


\subsubsection{IMF power-law index correction } \label{sec:result_binary}
\citet{Liu-2019MNRAS.490..550L} examined the binarity properties of field stars with masses ranging from 0.4 to 0.85 \M\ in the solar neighbourhood, including the binary fraction across various masses and metallicities. Additionally, \citet{Moe-2019ApJ...875...61M} studied the binary fraction as a function of metallicity for solar-type stars with primary masses ranging from 0.6 to 1.5 \M. Based on these studies, we roughly estimated the binary fraction for stars with mass $\leq$ 0.525 \M\ and mass > 0.525 \M\ in different \feh\ bins and derived the IMF-index corrections using the method of Section \ref{sec:simu_binaries}.

For the entire sample without [Fe/H] bin, the binary fractions of stars with mass $\leq$0.525 and > 0.525 \M\ are set to 24\% and 32\%, respectively. The corresponding shift values in the power-law index are 0.29 and 0.36, respectively. The corrected power-law indices are 1.48$\pm$0.03 ($\alpha_{\rm{1,corr}}$) and  2.17$\pm$0.03 ($\alpha_{\rm{2,corr}}$). They are in good agreement with those of Kroupa's IMF, where $\alpha_1$ and $\alpha_2$ are 1.3$\pm$0.5 and 2.3$\pm$0.3, respectively.

\begin{figure}
\centering
\includegraphics[width=0.5\textwidth, trim=0.0cm 1.0cm 0.0cm 0.0cm, clip] {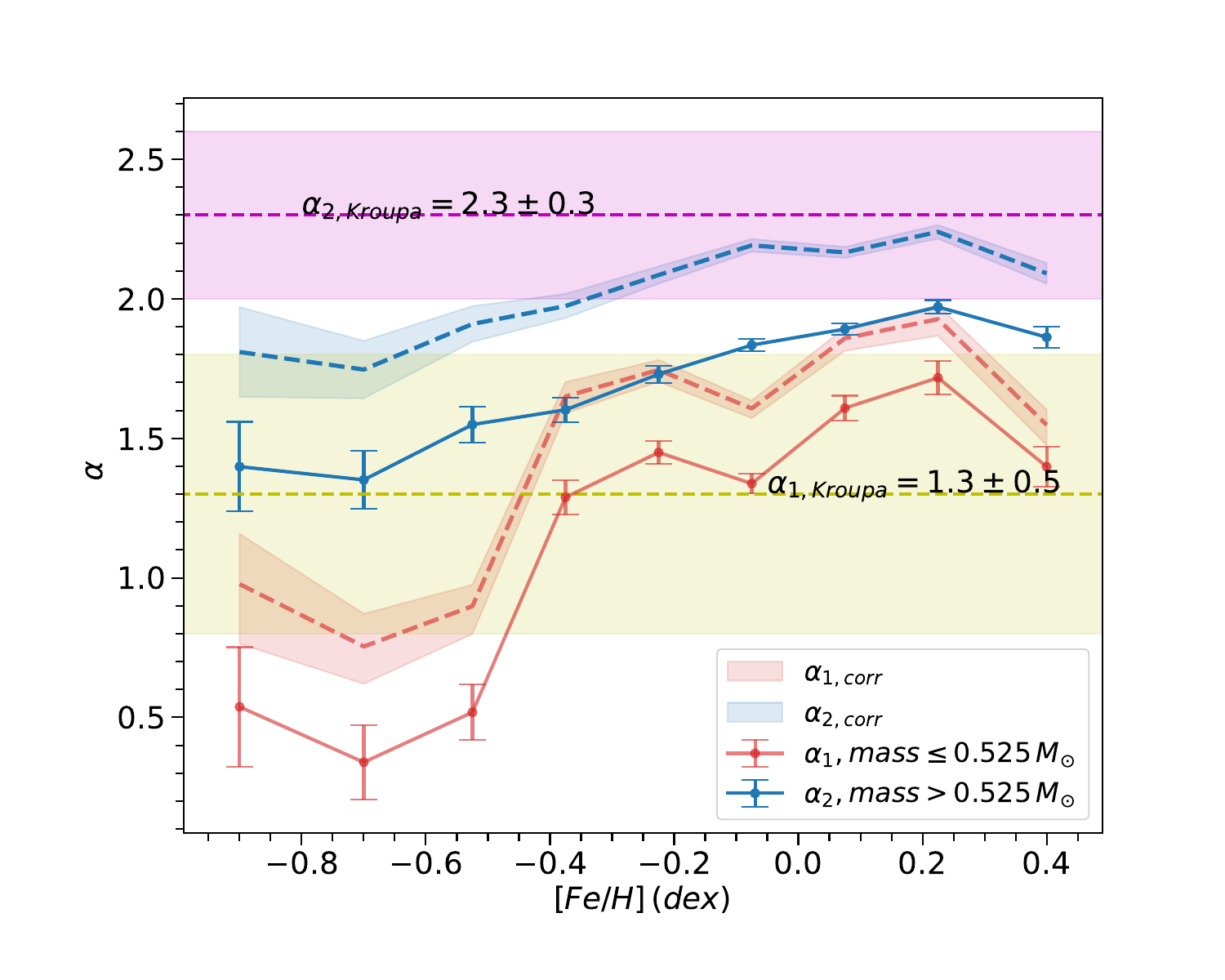}
\caption{The two solid lines are the same as those in Figure \ref{fig:alpha_feh}, while the two dashed lines represent the corresponding results after correcting for binarity. 
}\label{fig:alpha_cor}
\end{figure}

In Figure \ref{fig:alpha_cor}, the blue and red dashed lines show the corrected power-law indices of $\alpha_{\rm{1,corr}}$ and $\alpha_{\rm{2,corr}}$ versus \feh, respectively. The $\alpha_{\rm{1,corr}}$ changes from 0.97$\pm$0.21 to 1.55$\pm$0.07, and $\alpha_{\rm{2,corr}}$ varies from 1.78$\pm$0.16 to 2.09$\pm$0.04. It is obvious that the corrected indices still exhibit a statistically significant correlation with \feh. 

\subsection{IMF power-law index with different break points} \label{sec:beark_point}

\begin{figure*}
\centering
\includegraphics[width=1\textwidth, trim=0.0cm 0.0cm 0.0cm 0.0cm, clip] {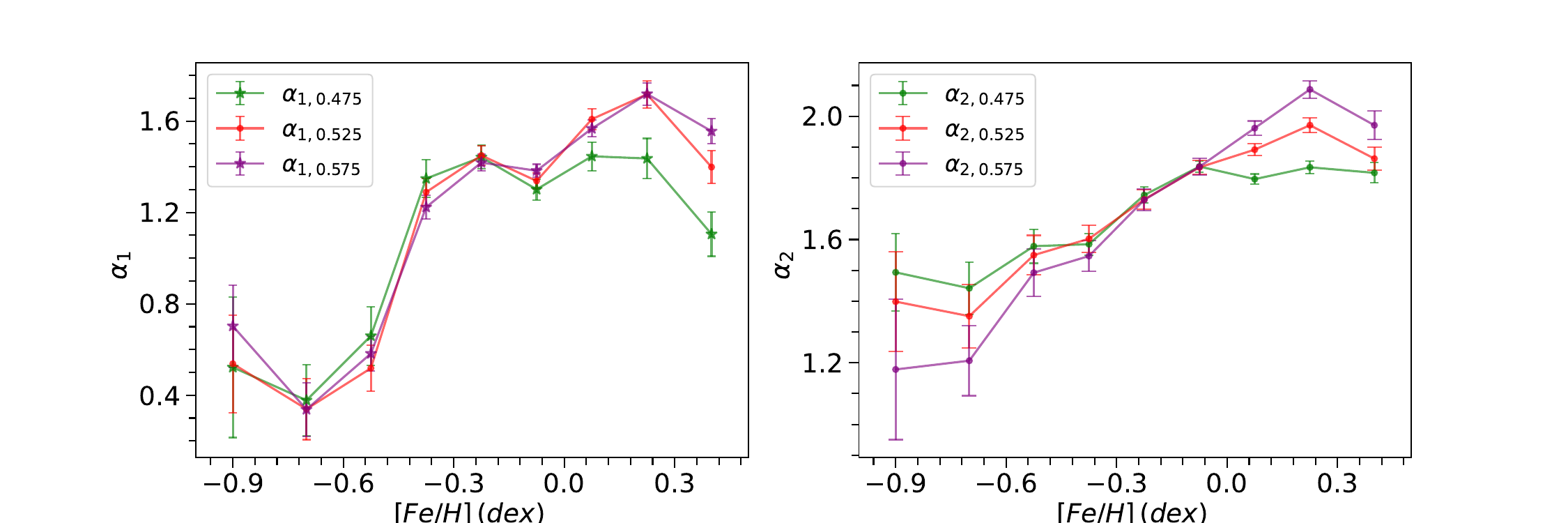}
\caption{The IMF power-law indices as a function of metallicity for three different break points. The left panel exhibits $\rm \alpha_{1,*}$ versus \feh, where * denotes the adopted break point: 0.475 (green), 0.525 (red), or 0.575 (purple). The right panel is the same as the left panel, but for the distribution of $\alpha_{2,*}$ and \feh. }\label{fig:breakpoint}
\end{figure*}

To test the sensitivity of our results to the adopted different break points. We repeated the power-law fit with break points at 0.475 and 0.575 \M\ in addition to the reference one of 0.525 \M . Using the method of Subsection \ref{sec:power_law}, we derived the corresponding low- and high-mass IMF power-law indices. Figure \ref{fig:breakpoint} exhibits the $\rm \alpha_{1,*}$ (left panel) and $\rm \alpha_{2,*}$ (right panel) versus \feh\ for all three break points, where * can be 0.475, 0.525, and 0.575. In every case, both indices increase with metallicity, indicating that the $\alpha$-\feh\ trend is statistically robust and only weakly sensitive to the exact choice of break point.

\subsection{The Stellar IMF derived from different mass bin size } \label{sec:mass_bin}

\begin{figure*}
\centering
\includegraphics[width=0.95\textwidth, trim=0.0cm 0.0cm 0.0cm 0.0cm, clip] {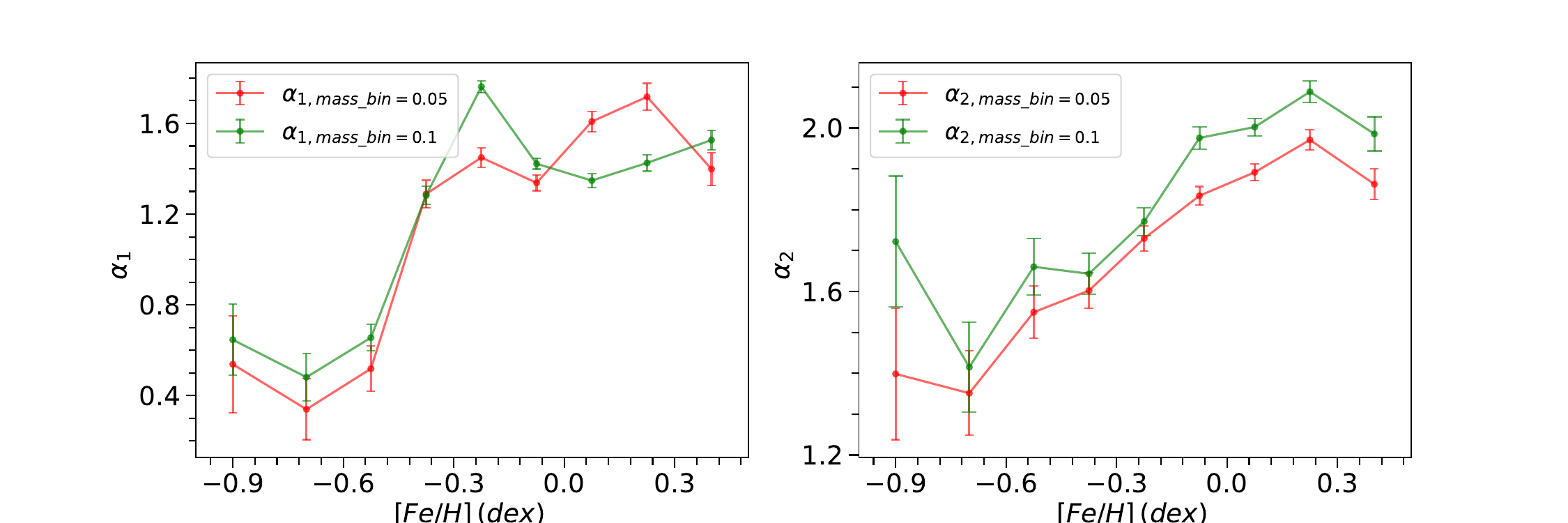}
\caption{The left panel shows $\alpha_1$ as a function of [Fe/H]. The red line represents $\alpha_1$
derived from IMFs constructed with a mass-bin size of 
0.05 \M, while the green line corresponds to a mass bin size of 0.10 \M. The right panel presents the same comparison, but for $\alpha_2$. }\label{fig:massbin}
\end{figure*}

To assess the impact of the mass-bin width on the IMF, we further explore the IMF using a bin width of 0.10 \M. We derived the $\alpha_1$ and $\alpha_2$ by setting the break point at 0.5\M. As shown in Figure \ref{fig:massbin}, the slopes of IMFs with mass bin sizes of 0.05 (red line) and 0.10 (green line) show similar trends with \feh. This indicates that our IMF results are insensitive to the choice of mass-bin size. 

\subsection{Caveat}
It should be noted that the SLAM predicted metallicities of M dwarfs with values below -0.6 dex may suffer from large uncertainty due to an insufficient training dataset, as mentioned in section \ref{subsec:metallicity}. Since most of these M dwarfs have masses below 0.525 \M , the uncertainties of $\alpha_1$ for stars with \feh\ < -0.6 dex may be larger than the nominal values.

\section{Conclusion} \label{sec:Conclusion}
We analysed over 500,000 LAMOST dwarf stars with masses between 0.25 and 1 \M\ and distances of 150-350 pc to investigate how the stellar IMF varies with metallicity. The \feh\ ranges from -1 to +0.5 dex. We trained an XGBoost model based on the PARSEC isochrones to predict the mass of each star with known $\rm M_{K_{s0}}$, \teff, and \feh. By splitting the whole sample into different mass and \feh\ bins, the intrinsic number density of stars was determined by correcting the observed selection function based on the 2MASS survey. We treated the vertical-integrated space density in each mass bin as proportional to the IMF.

We fitted the resulting IMFs with a broken power-law fixed at 0.525 \M. For the full sample, we obtained $\alpha_1$=1.19 $\pm$ 0.03 and $\alpha_2$=1.81$\pm$0.03. When the data were divided into metallicity bins, the results show that both IMF power-law indices increased systematically with \feh. These findings align with the variation trend reported by Li23. It suggests that a larger fraction of low-mass stars is formed in a metal-rich environment than in a metal-poor environment.


To investigate the impact of unresolved binaries on our IMF indices, we generated a mock data set comprising 300,000 single stars and quantified the binary‐induced shift in $\alpha$ as a function of the binary fraction for stars with mass $\leq$ 0.525 \M\ and > 0.525 \M. Based on the binary fractions reported by \citet{Moe-2019ApJ...875...61M} and \citet{Liu-2019MNRAS.490..550L}, we derived the binary fraction of our sample and applied the corresponding shift values to $\alpha_1$ and $\alpha_2$.

After correcting the effect of unresolved binaries, the aggregate sample (no metallicity binning) yields the adjusted values $ \alpha_{\rm{1,corr}} $ =
1.48$\pm$0.03 and $ \alpha_{\rm{2,corr}} $ = 2.17$\pm$0.03, fully consistent with the Kroupa's IMF. For the stars with different metallicities, $\alpha_{\rm{1,corr}}$ rises from 0.97 $\pm$ 0.21 at [Fe/H] = -1.0 dex to 1.55 $\pm$ 0.07 at 0.5 dex, 
while $\alpha_{\rm{2,corr}}$ increases from 1.78 $\pm$ 0.16 to 2.09$\pm$ 0.04. Thus both corrected indices keep increasing with [Fe/H].

We also tested alternative break points of 0.475 and 0.575 \M. For every break point, $\alpha_1$ and $\alpha_2$ maintain the increasing trend with \feh, demonstrating that the trend is insensitive to the exact transition mass. Finally, using mass-bin widths of 0.05 and 0.10 \M\ yields indistinguishable [Fe/H]–dependent trends in the inferred IMF slopes, implying that our results are insensitive to reasonable choices of mass-bin size.

Compared to LAMOST, the upcoming SDSS-V \citep{Almeida-2023} has the capability to detect fainter stars, making it a promising dataset to not only supplement the lack of metal-poor (<-0.6 dex) M dwarfs in FGK+M wide binaries but also provide lower mass stars (<0.25 \M). Studying IMFs of stars in the lower mass range is of great significance for understanding the properties of star formation.

\section*{Acknowledgements}
JL thanks the support from the European Research Council through ERC Advanced Grant No. 101054731.
BZ acknowledges the support from Natural Science Foundation of China (NSFC) under grant No.12203068. Guoshoujing Telescope (the Large Sky Area Multi-Object Fiber Spectroscopic Telescope LAMOST) is a National Major Scientific Project built by the Chinese Academy of Sciences. Funding for the project has been provided by the National Development and Reform Commission. LAMOST is operated and managed by the National Astronomical Observatories, Chinese Academy of Sciences. This work has made use of data from the European Space Agency (ESA) mission Gaia (https://www.cosmos.esa.int/gaia), processed by the Gaia Data Processing and Analysis Consortium (DPAC; https://www.cosmos.esa.int/web/gaia/dpac/consortium). Funding for the DPAC has been provided by national institutions, in particular the institutions participating in the Gaia Multilateral Agreement.

\section*{Data Availability}
The data used in Figures~\ref{fig:alpha_cor}, \ref{fig:breakpoint}, and \ref{fig:massbin} are available at \url{https://nadc.china-vo.org/res/r101699/}.



\bibliographystyle{mnras}
\bibliography{example} 

@ARTICLE{Salpeter-1955,
       author = {{Salpeter}, Edwin E.},
        title = "{The Luminosity Function and Stellar Evolution.}",
      journal = {\apj},
         year = 1955,
        month = jan,
       volume = {121},
        pages = {161},
          doi = {10.1086/145971},
       adsurl = {https://ui.adsabs.harvard.edu/abs/1955ApJ...121..161S}
}

@ARTICLE{Qiu-2024,
       author = {{Qiu}, Dan and {Li}, Jiadong and {Zhang}, Bo and {Liu}, Chao and {Tian}, Haijun and {Niu}, Zexi},
        title = "{Calibration of metallicity of LAMOST M dwarf stars using FGK+M wide binaries}",
      journal = {\mnras},
         year = 2024,
        month = feb,
       volume = {527},
       number = {4},
        pages = {11866-11881},
          doi = {10.1093/mnras/stad3950},
archivePrefix = {arXiv},
       eprint = {2312.12827},
 primaryClass = {astro-ph.SR},
       adsurl = {https://ui.adsabs.harvard.edu/abs/2024MNRAS.52711866Q}
}

@ARTICLE{Badry-2021,
       author = {{El-Badry}, Kareem and {Rix}, Hans-Walter and {Heintz}, Tyler M.},
        title = "{A million binaries from Gaia eDR3: sample selection and validation of Gaia parallax uncertainties}",
      journal = {\mnras},
         year = 2021,
        month = sep,
       volume = {506},
       number = {2},
        pages = {2269-2295},
          doi = {10.1093/mnras/stab323},
archivePrefix = {arXiv},
       eprint = {2101.05282},
 primaryClass = {astro-ph.SR},
       adsurl = {https://ui.adsabs.harvard.edu/abs/2021MNRAS.506.2269E}
}

@ARTICLE{Zhang-2020ApJS..246....9Z,
       author = {{Zhang}, Bo and {Liu}, Chao and {Deng}, Li-Cai},
        title = "{Deriving the Stellar Labels of LAMOST Spectra with the Stellar LAbel Machine (SLAM)}",
      journal = {\apjs},
         year = 2020,
        month = jan,
       volume = {246},
       number = {1},
          eid = {9},
        pages = {9},
          doi = {10.3847/1538-4365/ab55ef},
archivePrefix = {arXiv},
       eprint = {1908.08677},
 primaryClass = {astro-ph.SR},
       adsurl = {https://ui.adsabs.harvard.edu/abs/2020ApJS..246....9Z}
}

@ARTICLE{Bressan-2012MNRAS.427..127B,
       author = {{Bressan}, Alessandro and {Marigo}, Paola and {Girardi}, L{\'e}o. and {Salasnich}, Bernardo and {Dal Cero}, Claudia and {Rubele}, Stefano and {Nanni}, Ambra},
        title = "{PARSEC: stellar tracks and isochrones with the PAdova and TRieste Stellar Evolution Code}",
      journal = {\mnras},
         year = 2012,
        month = nov,
       volume = {427},
       number = {1},
        pages = {127-145},
          doi = {10.1111/j.1365-2966.2012.21948.x},
archivePrefix = {arXiv},
       eprint = {1208.4498},
 primaryClass = {astro-ph.SR},
       adsurl = {https://ui.adsabs.harvard.edu/abs/2012MNRAS.427..127B}
}

@ARTICLE{Chen-2014MNRAS.444.2525C,
       author = {{Chen}, Yang and {Girardi}, L{\'e}o and {Bressan}, Alessandro and {Marigo}, Paola and {Barbieri}, Mauro and {Kong}, Xu},
        title = "{Improving PARSEC models for very low mass stars}",
      journal = {\mnras},
         year = 2014,
        month = nov,
       volume = {444},
       number = {3},
        pages = {2525-2543},
          doi = {10.1093/mnras/stu1605},
archivePrefix = {arXiv},
       eprint = {1409.0322},
 primaryClass = {astro-ph.SR},
       adsurl = {https://ui.adsabs.harvard.edu/abs/2014MNRAS.444.2525C}
}

@ARTICLE{Chen-2016arXiv160302754C,
       author = {{Chen}, Tianqi and {Guestrin}, Carlos},
        title = "{XGBoost: A Scalable Tree Boosting System}",
      journal = {arXiv e-prints},
         year = 2016,
        month = mar,
          eid = {arXiv:1603.02754},
        pages = {arXiv:1603.02754},
          doi = {10.48550/arXiv.1603.02754},
archivePrefix = {arXiv},
       eprint = {1603.02754},
 primaryClass = {cs.LG},
       adsurl = {https://ui.adsabs.harvard.edu/abs/2016arXiv160302754C}
}

@ARTICLE{Birky-2020ApJ...892...31B,
       author = {{Birky}, Jessica and {Hogg}, David W. and {Mann}, Andrew W. and {Burgasser}, Adam},
        title = "{Temperatures and Metallicities of M Dwarfs in the APOGEE Survey}",
      journal = {\apj},
         year = 2020,
        month = mar,
       volume = {892},
       number = {1},
          eid = {31},
        pages = {31},
          doi = {10.3847/1538-4357/ab7004},
archivePrefix = {arXiv},
       eprint = {2001.04962},
 primaryClass = {astro-ph.SR},
       adsurl = {https://ui.adsabs.harvard.edu/abs/2020ApJ...892...31B}
}

@ARTICLE{Li-2023Natur.613..460L,
       author = {{Li}, Jiadong and {Liu}, Chao and {Zhang}, Zhi-Yu and {Tian}, Hao and {Fu}, Xiaoting and {Li}, Jiao and {Yan}, Zhi-Qiang},
        title = "{Stellar initial mass function varies with metallicity and time}",
      journal = {\nat},
         year = 2023,
        month = jan,
       volume = {613},
       number = {7944},
        pages = {460-462},
          doi = {10.1038/s41586-022-05488-1},
archivePrefix = {arXiv},
       eprint = {2301.07029},
 primaryClass = {astro-ph.GA},
       adsurl = {https://ui.adsabs.harvard.edu/abs/2023Natur.613..460L}
}

@ARTICLE{Mann-2019ApJ...871...63M,
       author = {{Mann}, Andrew W. and {Dupuy}, Trent and {Kraus}, Adam L. and {Gaidos}, Eric and {Ansdell}, Megan and {Ireland}, Michael and {Rizzuto}, Aaron C. and {Hung}, Chao-Ling and {Dittmann}, Jason and {Factor}, Samuel and {Feiden}, Gregory and {Martinez}, Raquel A. and {Ru{\'\i}z-Rodr{\'\i}guez}, Dary and {Thao}, Pa Chia},
        title = "{How to Constrain Your M Dwarf. II. The Mass-Luminosity-Metallicity Relation from 0.075 to 0.70 Solar Masses}",
      journal = {\apj},
         year = 2019,
        month = jan,
       volume = {871},
       number = {1},
          eid = {63},
        pages = {63},
          doi = {10.3847/1538-4357/aaf3bc},
archivePrefix = {arXiv},
       eprint = {1811.06938},
 primaryClass = {astro-ph.SR},
       adsurl = {https://ui.adsabs.harvard.edu/abs/2019ApJ...871...63M}
}

@ARTICLE{Liu-2017RAA....17...96L,
       author = {{Liu}, Chao and {Xu}, Yan and {Wan}, Jun-Chen and {Wang}, Hai-Feng and {Carlin}, Jeffrey L. and {Deng}, Li-Cai and {Newberg}, Heidi Jo and {Cao}, Zi-Huang and {Hou}, Yong-Hui and {Wang}, Yue-Fei and {Zhang}, Yong},
        title = "{Mapping the Milky Way with LAMOST I: method and overview}",
      journal = {Research in Astronomy and Astrophysics},
         year = 2017,
        month = sep,
       volume = {17},
       number = {9},
          eid = {096},
        pages = {096},
          doi = {10.1088/1674-4527/17/9/96},
archivePrefix = {arXiv},
       eprint = {1701.07831},
 primaryClass = {astro-ph.GA},
       adsurl = {https://ui.adsabs.harvard.edu/abs/2017RAA....17...96L}
}

@ARTICLE{Skrutskie-2006AJ....131.1163S,
       author = {{Skrutskie}, M.~F. and {Cutri}, R.~M. and {Stiening}, R. and {Weinberg}, M.~D. and {Schneider}, S. and {Carpenter}, J.~M. and {Beichman}, C. and {Capps}, R. and {Chester}, T. and {Elias}, J. and {Huchra}, J. and {Liebert}, J. and {Lonsdale}, C. and {Monet}, D.~G. and {Price}, S. and {Seitzer}, P. and {Jarrett}, T. and {Kirkpatrick}, J.~D. and {Gizis}, J.~E. and {Howard}, E. and {Evans}, T. and {Fowler}, J. and {Fullmer}, L. and {Hurt}, R. and {Light}, R. and {Kopan}, E.~L. and {Marsh}, K.~A. and {McCallon}, H.~L. and {Tam}, R. and {Van Dyk}, S. and {Wheelock}, S.},
        title = "{The Two Micron All Sky Survey (2MASS)}",
      journal = {\aj},
         year = 2006,
        month = feb,
       volume = {131},
       number = {2},
        pages = {1163-1183},
          doi = {10.1086/498708},
       adsurl = {https://ui.adsabs.harvard.edu/abs/2006AJ....131.1163S}
}

@ARTICLE{Green-2019ApJ...887...93G,
       author = {{Green}, Gregory M. and {Schlafly}, Edward and {Zucker}, Catherine and {Speagle}, Joshua S. and {Finkbeiner}, Douglas},
        title = "{A 3D Dust Map Based on Gaia, Pan-STARRS 1, and 2MASS}",
      journal = {\apj},
         year = 2019,
        month = dec,
       volume = {887},
       number = {1},
          eid = {93},
        pages = {93},
          doi = {10.3847/1538-4357/ab5362},
archivePrefix = {arXiv},
       eprint = {1905.02734},
 primaryClass = {astro-ph.GA},
       adsurl = {https://ui.adsabs.harvard.edu/abs/2019ApJ...887...93G}
}

@ARTICLE{Wang-2019ApJ...877..116W,
       author = {{Wang}, Shu and {Chen}, Xiaodian},
        title = "{The Optical to Mid-infrared Extinction Law Based on the APOGEE, Gaia DR2, Pan-STARRS1, SDSS, APASS, 2MASS, and WISE Surveys}",
      journal = {\apj},
         year = 2019,
        month = jun,
       volume = {877},
       number = {2},
          eid = {116},
        pages = {116},
          doi = {10.3847/1538-4357/ab1c61},
archivePrefix = {arXiv},
       eprint = {1904.04575},
 primaryClass = {astro-ph.GA},
       adsurl = {https://ui.adsabs.harvard.edu/abs/2019ApJ...877..116W}
}

@ARTICLE{2015MNRAS.447.3526K,
       author = {{Kordopatis}, G. and {Binney}, J. and {Gilmore}, G. and {Wyse}, R.~F.~G. and {Belokurov}, V. and {McMillan}, P.~J. and {Hatfield}, P. and {Grebel}, E.~K. and {Steinmetz}, M. and {Navarro}, J.~F. and {Seabroke}, G. and {Minchev}, I. and {Chiappini}, C. and {Bienaym{\'e}}, O. and {Bland-Hawthorn}, J. and {Freeman}, K.~C. and {Gibson}, B.~K. and {Helmi}, A. and {Munari}, U. and {Parker}, Q. and {Reid}, W.~A. and {Siebert}, A. and {Siviero}, A. and {Zwitter}, T.},
        title = "{The rich are different: evidence from the RAVE survey for stellar radial migration}",
      journal = {\mnras},
         year = 2015,
        month = mar,
       volume = {447},
       number = {4},
        pages = {3526-3535},
          doi = {10.1093/mnras/stu2726},
archivePrefix = {arXiv},
       eprint = {1412.5649},
 primaryClass = {astro-ph.GA},
       adsurl = {https://ui.adsabs.harvard.edu/abs/2015MNRAS.447.3526K}
}

@ARTICLE{Kroupa-2001MNRAS.322..231K,
       author = {{Kroupa}, Pavel},
        title = "{On the variation of the initial mass function}",
      journal = {\mnras},
         year = 2001,
        month = apr,
       volume = {322},
       number = {2},
        pages = {231-246},
          doi = {10.1046/j.1365-8711.2001.04022.x},
archivePrefix = {arXiv},
       eprint = {astro-ph/0009005},
 primaryClass = {astro-ph},
       adsurl = {https://ui.adsabs.harvard.edu/abs/2001MNRAS.322..231K}
}

@ARTICLE{Liu-2019MNRAS.490..550L,
       author = {{Liu}, Chao},
        title = "{Smoking gun of the dynamical processing of solar-type field binary stars}",
      journal = {\mnras},
         year = 2019,
        month = nov,
       volume = {490},
       number = {1},
        pages = {550-565},
          doi = {10.1093/mnras/stz2274},
archivePrefix = {arXiv},
       eprint = {1907.02250},
 primaryClass = {astro-ph.SR},
       adsurl = {https://ui.adsabs.harvard.edu/abs/2019MNRAS.490..550L}
}

@ARTICLE{Moe-2019ApJ...875...61M,
       author = {{Moe}, Maxwell and {Kratter}, Kaitlin M. and {Badenes}, Carles},
        title = "{The Close Binary Fraction of Solar-type Stars Is Strongly Anticorrelated with Metallicity}",
      journal = {\apj},
         year = 2019,
        month = apr,
       volume = {875},
       number = {1},
          eid = {61},
        pages = {61},
          doi = {10.3847/1538-4357/ab0d88},
archivePrefix = {arXiv},
       eprint = {1808.02116},
 primaryClass = {astro-ph.SR},
       adsurl = {https://ui.adsabs.harvard.edu/abs/2019ApJ...875...61M}
}

@ARTICLE{Ma-2005ApJ...629..873M,
       author = {{Ma{\'\i}z Apell{\'a}niz}, J. and {{\'U}beda}, L.},
        title = "{Numerical Biases on Initial Mass Function Determinations Created by Binning}",
      journal = {\apj},
         year = 2005,
        month = aug,
       volume = {629},
       number = {2},
        pages = {873-880},
          doi = {10.1086/431458},
archivePrefix = {arXiv},
       eprint = {astro-ph/0505012},
 primaryClass = {astro-ph},
       adsurl = {https://ui.adsabs.harvard.edu/abs/2005ApJ...629..873M}
}

@ARTICLE{Cara-2008ApJ...686..148C,
       author = {{Cara}, M. and {Lister}, M.~L.},
        title = "{Avoiding Spurious Breaks in Binned Luminosity Functions}",
      journal = {\apj},
         year = 2008,
        month = oct,
       volume = {686},
       number = {1},
        pages = {148-154},
          doi = {10.1086/590902},
archivePrefix = {arXiv},
       eprint = {0806.1232},
 primaryClass = {astro-ph},
       adsurl = {https://ui.adsabs.harvard.edu/abs/2008ApJ...686..148C}
}

@ARTICLE{Je-2018A&A...620A..39J,
       author = {{Je{\v{r}}{\'a}bkov{\'a}}, T. and {Hasani Zonoozi}, A. and {Kroupa}, P. and {Beccari}, G. and {Yan}, Z. and {Vazdekis}, A. and {Zhang}, Z. -Y.},
        title = "{Impact of metallicity and star formation rate on the time-dependent, galaxy-wide stellar initial mass function}",
      journal = {\aap},
         year = 2018,
        month = nov,
       volume = {620},
          eid = {A39},
        pages = {A39},
          doi = {10.1051/0004-6361/201833055},
archivePrefix = {arXiv},
       eprint = {1809.04603},
 primaryClass = {astro-ph.GA},
       adsurl = {https://ui.adsabs.harvard.edu/abs/2018A&A...620A..39J}
}

@ARTICLE{Dabringhausen-2009MNRAS.394.1529D,
       author = {{Dabringhausen}, J. and {Kroupa}, P. and {Baumgardt}, H.},
        title = "{A top-heavy stellar initial mass function in starbursts as an explanation for the high mass-to-light ratios of ultra-compact dwarf galaxies}",
      journal = {\mnras},
         year = 2009,
        month = apr,
       volume = {394},
       number = {3},
        pages = {1529-1543},
          doi = {10.1111/j.1365-2966.2009.14425.x},
archivePrefix = {arXiv},
       eprint = {0901.0915},
 primaryClass = {astro-ph.GA},
       adsurl = {https://ui.adsabs.harvard.edu/abs/2009MNRAS.394.1529D}
}

@ARTICLE{Kalari-2018ApJ...857..132K,
       author = {{Kalari}, Venu M. and {Carraro}, Giovanni and {Evans}, Christopher J. and {Rubio}, Monica},
        title = "{The Magellanic Bridge Cluster NGC 796: Deep Optical AO Imaging Reveals the Stellar Content and Initial Mass Function of a Massive Open Cluster}",
      journal = {\apj},
         year = 2018,
        month = apr,
       volume = {857},
       number = {2},
          eid = {132},
        pages = {132},
          doi = {10.3847/1538-4357/aab609},
archivePrefix = {arXiv},
       eprint = {1801.01490},
 primaryClass = {astro-ph.SR},
       adsurl = {https://ui.adsabs.harvard.edu/abs/2018ApJ...857..132K}
}

@ARTICLE{Elmegreen-2004MNRAS.354..367E,
       author = {{Elmegreen}, B.~G.},
        title = "{Variability in the stellar initial mass function at low and high mass: three-component IMF models}",
      journal = {\mnras},
         year = 2004,
        month = oct,
       volume = {354},
       number = {2},
        pages = {367-374},
          doi = {10.1111/j.1365-2966.2004.08187.x},
archivePrefix = {arXiv},
       eprint = {astro-ph/0408231},
 primaryClass = {astro-ph},
       adsurl = {https://ui.adsabs.harvard.edu/abs/2004MNRAS.354..367E}
}

@ARTICLE{Weidner-2005ApJ...625..754W,
       author = {{Weidner}, C. and {Kroupa}, P.},
        title = "{The Variation of Integrated Star Initial Mass Functions among Galaxies}",
      journal = {\apj},
         year = 2005,
        month = jun,
       volume = {625},
       number = {2},
        pages = {754-762},
          doi = {10.1086/429867},
archivePrefix = {arXiv},
       eprint = {astro-ph/0502525},
 primaryClass = {astro-ph},
       adsurl = {https://ui.adsabs.harvard.edu/abs/2005ApJ...625..754W}
}

@PHDTHESIS{Levine-2006PhDT........29L,
       author = {{Levine}, Joanna Lisa},
        title = "{Low mass star and brown dwarf formation in the Orion B molecular cloud}",
       school = {University of Florida},
         year = 2006,
        month = jan,
       adsurl = {https://ui.adsabs.harvard.edu/abs/2006PhDT........29L}
}

@ARTICLE{Larson-2005MNRAS.359..211L,
       author = {{Larson}, Richard B.},
        title = "{Thermal physics, cloud geometry and the stellar initial mass function}",
      journal = {\mnras},
         year = 2005,
        month = may,
       volume = {359},
       number = {1},
        pages = {211-222},
          doi = {10.1111/j.1365-2966.2005.08881.x},
archivePrefix = {arXiv},
       eprint = {astro-ph/0412357},
 primaryClass = {astro-ph},
       adsurl = {https://ui.adsabs.harvard.edu/abs/2005MNRAS.359..211L}
}

@ARTICLE{Elmegreen-2008ApJ...681..365E,
       author = {{Elmegreen}, Bruce G. and {Klessen}, Ralf S. and {Wilson}, Christine D.},
        title = "{On the Constancy of the Characteristic Mass of Young Stars}",
      journal = {\apj},
         year = 2008,
        month = jul,
       volume = {681},
       number = {1},
        pages = {365-374},
          doi = {10.1086/588725},
archivePrefix = {arXiv},
       eprint = {0803.4411},
 primaryClass = {astro-ph},
       adsurl = {https://ui.adsabs.harvard.edu/abs/2008ApJ...681..365E}
}

@INPROCEEDINGS{Dabringhausen-2011ASPC..440..261D,
       author = {{Dabringhausen}, J. and {Kroupa}, P.},
        title = "{Top-heavy Initial Mass Functions in Ultra-compact Dwarf Galaxies?}",
    booktitle = {UP2010: Have Observations Revealed a Variable Upper End of the Initial Mass Function?},
         year = 2011,
       editor = {{Treyer}, M. and {Wyder}, T. and {Neill}, J. and {Seibert}, M. and {Lee}, J.},
       series = {Astronomical Society of the Pacific Conference Series},
       volume = {440},
        month = jun,
        pages = {261},
          doi = {10.48550/arXiv.1201.3912},
archivePrefix = {arXiv},
       eprint = {1201.3912},
 primaryClass = {astro-ph.CO},
       adsurl = {https://ui.adsabs.harvard.edu/abs/2011ASPC..440..261D}
}

@ARTICLE{Cescutti-2011A&A...525A.126C,
       author = {{Cescutti}, G. and {Matteucci}, F.},
        title = "{Galactic astroarchaeology: reconstructing the bulge history by means of the newest data}",
      journal = {\aap},
         year = 2011,
        month = jan,
       volume = {525},
          eid = {A126},
        pages = {A126},
          doi = {10.1051/0004-6361/201015665},
archivePrefix = {arXiv},
       eprint = {1010.1469},
 primaryClass = {astro-ph.GA},
       adsurl = {https://ui.adsabs.harvard.edu/abs/2011A&A...525A.126C}
}

@INPROCEEDINGS{Kobayashi-2010AIPC.1240..123K,
       author = {{Kobayashi}, Chiaki},
        title = "{Chemodynamical Simulations with Variable IMF}",
    booktitle = {Hunting for the Dark: the Hidden Side of Galaxy Formation},
         year = 2010,
       editor = {{Debattista}, Victor P. and {Popescu}, C.~C.},
       series = {American Institute of Physics Conference Series},
       volume = {1240},
        month = jun,
    publisher = {AIP},
        pages = {123-126},
          doi = {10.1063/1.3458465},
archivePrefix = {arXiv},
       eprint = {1002.4475},
 primaryClass = {astro-ph.CO},
       adsurl = {https://ui.adsabs.harvard.edu/abs/2010AIPC.1240..123K}
}

@ARTICLE{Spiniello-2012ApJ...753L..32S,
       author = {{Spiniello}, C. and {Trager}, S.~C. and {Koopmans}, L.~V.~E. and {Chen}, Y.~P.},
        title = "{Evidence for a Mild Steepening and Bottom-heavy Initial Mass Function in Massive Galaxies from Sodium and Titanium-oxide Indicators}",
      journal = {\apjl},
         year = 2012,
        month = jul,
       volume = {753},
       number = {2},
          eid = {L32},
        pages = {L32},
          doi = {10.1088/2041-8205/753/2/L32},
archivePrefix = {arXiv},
       eprint = {1204.3823},
 primaryClass = {astro-ph.CO},
       adsurl = {https://ui.adsabs.harvard.edu/abs/2012ApJ...753L..32S}
}

@ARTICLE{Marks-2012MNRAS.422.2246M,
       author = {{Marks}, Michael and {Kroupa}, Pavel and {Dabringhausen}, J{\"o}rg and {Pawlowski}, Marcel S.},
        title = "{Evidence for top-heavy stellar initial mass functions with increasing density and decreasing metallicity}",
      journal = {\mnras},
         year = 2012,
        month = may,
       volume = {422},
       number = {3},
        pages = {2246-2254},
          doi = {10.1111/j.1365-2966.2012.20767.x},
archivePrefix = {arXiv},
       eprint = {1202.4755},
 primaryClass = {astro-ph.GA},
       adsurl = {https://ui.adsabs.harvard.edu/abs/2012MNRAS.422.2246M}
}

@ARTICLE{Geha-2013ApJ...771...29G,
       author = {{Geha}, Marla and {Brown}, Thomas M. and {Tumlinson}, Jason and {Kalirai}, Jason S. and {Simon}, Joshua D. and {Kirby}, Evan N. and {VandenBerg}, Don A. and {Mu{\~n}oz}, Ricardo R. and {Avila}, Roberto J. and {Guhathakurta}, Puragra and {Ferguson}, Henry C.},
        title = "{The Stellar Initial Mass Function of Ultra-faint Dwarf Galaxies: Evidence for IMF Variations with Galactic Environment}",
      journal = {\apj},
         year = 2013,
        month = jul,
       volume = {771},
       number = {1},
          eid = {29},
        pages = {29},
          doi = {10.1088/0004-637X/771/1/29},
archivePrefix = {arXiv},
       eprint = {1304.7769},
 primaryClass = {astro-ph.CO},
       adsurl = {https://ui.adsabs.harvard.edu/abs/2013ApJ...771...29G}
}

@INCOLLECTION{Kroupa-2013pss5.book..115K,
       author = {{Kroupa}, Pavel and {Weidner}, Carsten and {Pflamm-Altenburg}, Jan and {Thies}, Ingo and {Dabringhausen}, J{\"o}rg and {Marks}, Michael and {Maschberger}, Thomas},
        title = "{The Stellar and Sub-Stellar Initial Mass Function of Simple and Composite Populations}",
    booktitle = {Planets, Stars and Stellar Systems. Volume 5: Galactic Structure and Stellar Populations},
         year = 2013,
       editor = {{Oswalt}, Terry D. and {Gilmore}, Gerard},
       volume = {5},
        pages = {115},
          doi = {10.1007/978-94-007-5612-0_4},
       adsurl = {https://ui.adsabs.harvard.edu/abs/2013pss5.book..115K}
}

@ARTICLE{Conroy-2012ApJ...760...71C,
       author = {{Conroy}, Charlie and {van Dokkum}, Pieter G.},
        title = "{The Stellar Initial Mass Function in Early-type Galaxies From Absorption Line Spectroscopy. II. Results}",
      journal = {\apj},
         year = 2012,
        month = nov,
       volume = {760},
       number = {1},
          eid = {71},
        pages = {71},
          doi = {10.1088/0004-637X/760/1/71},
archivePrefix = {arXiv},
       eprint = {1205.6473},
 primaryClass = {astro-ph.CO},
       adsurl = {https://ui.adsabs.harvard.edu/abs/2012ApJ...760...71C}
}

@ARTICLE{van-2012ApJ...760...70V,
       author = {{van Dokkum}, Pieter G. and {Conroy}, Charlie},
        title = "{The Stellar Initial Mass Function in Early-type Galaxies from Absorption Line Spectroscopy. I. Data and Empirical Trends}",
      journal = {\apj},
         year = 2012,
        month = nov,
       volume = {760},
       number = {1},
          eid = {70},
        pages = {70},
          doi = {10.1088/0004-637X/760/1/70},
archivePrefix = {arXiv},
       eprint = {1205.6471},
 primaryClass = {astro-ph.CO},
       adsurl = {https://ui.adsabs.harvard.edu/abs/2012ApJ...760...70V}
}

@ARTICLE{Weidner-2013MNRAS.436.3309W,
       author = {{Weidner}, C. and {Kroupa}, P. and {Pflamm-Altenburg}, J. and {Vazdekis}, A.},
        title = "{The galaxy-wide initial mass function of dwarf late-type to massive early-type galaxies}",
      journal = {\mnras},
         year = 2013,
        month = dec,
       volume = {436},
       number = {4},
        pages = {3309-3320},
          doi = {10.1093/mnras/stt1806},
archivePrefix = {arXiv},
       eprint = {1309.6634},
 primaryClass = {astro-ph.CO},
       adsurl = {https://ui.adsabs.harvard.edu/abs/2013MNRAS.436.3309W}
}

@MISC{Adams-2013hst..prop13232A,
       author = {{Adams}, Joshua},
        title = "{Main Sequence Star Counts as a Probe of IMF Variations with Galactic Environment}",
 howpublished = {HST Proposal ID 13232. Cycle 21},
         year = 2013,
        month = oct,
        pages = {13232},
       adsurl = {https://ui.adsabs.harvard.edu/abs/2013hst..prop13232A}
}

@ARTICLE{Bekki-2013ApJ...779....9B,
       author = {{Bekki}, Kenji},
        title = "{Origin of a Bottom-heavy Stellar Initial Mass Function in Elliptical Galaxies}",
      journal = {\apj},
         year = 2013,
        month = dec,
       volume = {779},
       number = {1},
          eid = {9},
        pages = {9},
          doi = {10.1088/0004-637X/779/1/9},
archivePrefix = {arXiv},
       eprint = {1311.1633},
 primaryClass = {astro-ph.CO},
       adsurl = {https://ui.adsabs.harvard.edu/abs/2013ApJ...779....9B}
}

@INPROCEEDINGS{Spiniello-2016ASSP...42..219S,
       author = {{Spiniello}, C.},
        title = "{The Low-Mass End of the Initial Mass Function in Massive Early-Type-Galaxies}",
    booktitle = {The Universe of Digital Sky Surveys},
         year = 2016,
       editor = {{Napolitano}, Nicola R. and {Longo}, Giuseppe and {Marconi}, Marcella and {Paolillo}, Maurizio and {Iodice}, Enrichetta},
       series = {Astrophysics and Space Science Proceedings},
       volume = {42},
        month = jan,
        pages = {219},
          doi = {10.1007/978-3-319-19330-4_34},
       adsurl = {https://ui.adsabs.harvard.edu/abs/2016ASSP...42..219S}
}

@ARTICLE{Clauwens-2016MNRAS.462.2832C,
       author = {{Clauwens}, Bart and {Schaye}, Joop and {Franx}, Marijn},
        title = "{Implications of a variable IMF for the interpretation of observations of galaxy populations}",
      journal = {\mnras},
         year = 2016,
        month = nov,
       volume = {462},
       number = {3},
        pages = {2832-2846},
          doi = {10.1093/mnras/stw1808},
archivePrefix = {arXiv},
       eprint = {1603.05281},
 primaryClass = {astro-ph.GA},
       adsurl = {https://ui.adsabs.harvard.edu/abs/2016MNRAS.462.2832C}
}

@ARTICLE{Kroupa-2018arXiv180610605K,
       author = {{Kroupa}, Pavel and {Jerabkova}, Tereza},
        title = "{The Impact of Binaries on the Stellar Initial Mass Function}",
      journal = {arXiv e-prints},
         year = 2018,
        month = jun,
          eid = {arXiv:1806.10605},
        pages = {arXiv:1806.10605},
          doi = {10.48550/arXiv.1806.10605},
archivePrefix = {arXiv},
       eprint = {1806.10605},
 primaryClass = {astro-ph.GA},
       adsurl = {https://ui.adsabs.harvard.edu/abs/2018arXiv180610605K}
}

@ARTICLE{Gennaro-2018ApJ...855...20G,
       author = {{Gennaro}, Mario and {Tchernyshyov}, Kirill and {Brown}, Thomas M. and {Geha}, Marla and {Avila}, Roberto J. and {Guhathakurta}, Puragra and {Kalirai}, Jason S. and {Kirby}, Evan N. and {Renzini}, Alvio and {Simon}, Joshua D. and {Tumlinson}, Jason and {Vargas}, Luis C.},
        title = "{Evidence of a Non-universal Stellar Initial Mass Function. Insights from HST Optical Imaging of Six Ultra-faint Dwarf Milky Way Satellites}",
      journal = {\apj},
         year = 2018,
        month = mar,
       volume = {855},
       number = {1},
          eid = {20},
        pages = {20},
          doi = {10.3847/1538-4357/aaa973},
archivePrefix = {arXiv},
       eprint = {1801.06195},
 primaryClass = {astro-ph.GA},
       adsurl = {https://ui.adsabs.harvard.edu/abs/2018ApJ...855...20G}
}

@ARTICLE{Villaume-2017ApJ...850L..14V,
       author = {{Villaume}, Alexa and {Brodie}, Jean and {Conroy}, Charlie and {Romanowsky}, Aaron J. and {van Dokkum}, Pieter},
        title = "{Initial Mass Function Variability (or Not) among Low-velocity Dispersion, Compact Stellar Systems}",
      journal = {\apjl},
         year = 2017,
        month = nov,
       volume = {850},
       number = {1},
          eid = {L14},
        pages = {L14},
          doi = {10.3847/2041-8213/aa970f},
archivePrefix = {arXiv},
       eprint = {1710.11144},
 primaryClass = {astro-ph.GA},
       adsurl = {https://ui.adsabs.harvard.edu/abs/2017ApJ...850L..14V}
}

@ARTICLE{Lagattuta-2017ApJ...846..166L,
       author = {{Lagattuta}, David J. and {Mould}, Jeremy R. and {Forbes}, Duncan A. and {Monson}, Andrew J. and {Pastorello}, Nicola and {Persson}, S. Eric},
        title = "{Evidence of a Bottom-heavy Initial Mass Function in Massive Early-type Galaxies from Near-infrared Metal Lines}",
      journal = {\apj},
         year = 2017,
        month = sep,
       volume = {846},
       number = {2},
          eid = {166},
        pages = {166},
          doi = {10.3847/1538-4357/aa8563},
archivePrefix = {arXiv},
       eprint = {1708.04621},
 primaryClass = {astro-ph.GA},
       adsurl = {https://ui.adsabs.harvard.edu/abs/2017ApJ...846..166L}
}

@ARTICLE{Meyer-2019ApJ...875..151M,
       author = {{Meyer}, R. Elliot and {Sivanandam}, Suresh and {Moon}, Dae-Sik},
        title = "{Initial Mass Function Variation in Two Elliptical Galaxies Using Near-infrared Tracers}",
      journal = {\apj},
         year = 2019,
        month = apr,
       volume = {875},
       number = {2},
          eid = {151},
        pages = {151},
          doi = {10.3847/1538-4357/ab11d2},
archivePrefix = {arXiv},
       eprint = {1903.08323},
 primaryClass = {astro-ph.GA},
       adsurl = {https://ui.adsabs.harvard.edu/abs/2019ApJ...875..151M}
}

@ARTICLE{Barber-2019MNRAS.483..985B,
       author = {{Barber}, Christopher and {Schaye}, Joop and {Crain}, Robert A.},
        title = "{Calibrated, cosmological hydrodynamical simulations with variable IMFs III: spatially resolved properties and evolution}",
      journal = {\mnras},
         year = 2019,
        month = feb,
       volume = {483},
       number = {1},
        pages = {985-1002},
          doi = {10.1093/mnras/sty3011},
archivePrefix = {arXiv},
       eprint = {1807.11310},
 primaryClass = {astro-ph.GA},
       adsurl = {https://ui.adsabs.harvard.edu/abs/2019MNRAS.483..985B}
}

@ARTICLE{Hallakoun-2021MNRAS.507..398H,
       author = {{Hallakoun}, Na'ama and {Maoz}, Dan},
        title = "{A bottom-heavy initial mass function for the likely-accreted blue-halo stars of the Milky Way}",
      journal = {\mnras},
         year = 2021,
        month = oct,
       volume = {507},
       number = {1},
        pages = {398-413},
          doi = {10.1093/mnras/stab2145},
archivePrefix = {arXiv},
       eprint = {2009.05047},
 primaryClass = {astro-ph.GA},
       adsurl = {https://ui.adsabs.harvard.edu/abs/2021MNRAS.507..398H}
}

@ARTICLE{Dickson-2023MNRAS.522.5320D,
       author = {{Dickson}, N. and {H{\'e}nault-Brunet}, V. and {Baumgardt}, H. and {Gieles}, M. and {Smith}, P.~J.},
        title = "{Multimass modelling of Milky Way globular clusters - I. Implications on their stellar initial mass function above 1 M$_{{\ensuremath{\odot}}}$}",
      journal = {\mnras},
         year = 2023,
        month = jul,
       volume = {522},
       number = {4},
        pages = {5320-5339},
          doi = {10.1093/mnras/stad1254},
archivePrefix = {arXiv},
       eprint = {2303.01637},
 primaryClass = {astro-ph.GA},
       adsurl = {https://ui.adsabs.harvard.edu/abs/2023MNRAS.522.5320D}
}

@ARTICLE{Sharda-2023MNRAS.518.3985S,
       author = {{Sharda}, Piyush and {Amarsi}, Anish M. and {Grasha}, Kathryn and {Krumholz}, Mark R. and {Yong}, David and {Chiaki}, Gen and {Roy}, Arpita and {Nordlander}, Thomas},
        title = "{The impact of carbon and oxygen abundances on the metal-poor initial mass function}",
      journal = {\mnras},
         year = 2023,
        month = jan,
       volume = {518},
       number = {3},
        pages = {3985-3998},
          doi = {10.1093/mnras/stac3315},
archivePrefix = {arXiv},
       eprint = {2211.05505},
 primaryClass = {astro-ph.GA},
       adsurl = {https://ui.adsabs.harvard.edu/abs/2023MNRAS.518.3985S}
}

@ARTICLE{Yan-2024ApJ...969...95Y,
       author = {{Yan}, Zhiqiang and {Li}, Jiadong and {Kroupa}, Pavel and {Jerabkova}, Tereza and {Gjergo}, Eda and {Zhang}, Zhi-Yu},
        title = "{The Variation in the Galaxy-wide Initial Mass Function for Low-mass Stars: Modeling and Observational Insights}",
      journal = {\apj},
         year = 2024,
        month = jul,
       volume = {969},
       number = {2},
          eid = {95},
        pages = {95},
          doi = {10.3847/1538-4357/ad499d},
archivePrefix = {arXiv},
       eprint = {2405.05308},
 primaryClass = {astro-ph.GA},
       adsurl = {https://ui.adsabs.harvard.edu/abs/2024ApJ...969...95Y}
}

@ARTICLE{Maksymowicz-2024MNRAS.531.2864M,
       author = {{Maksymowicz-Maciata}, Michalina and {Spiniello}, Chiara and {Mart{\'\i}n-Navarro}, Ignacio and {Ferr{\'e}-Mateu}, Anna and {Bevacqua}, Davide and {Cappellari}, Michele and {D'Ago}, Giuseppe and {Tortora}, Crescenzo and {Arnaboldi}, Magda and {Hartke}, Johanna and {Napolitano}, Nicola R. and {Saracco}, Paolo and {Scognamiglio}, Diana},
        title = "{INSPIRE: INvestigating Stellar Population In RElics - VI. The low-mass end slope of the stellar initial mass function and chemical composition}",
      journal = {\mnras},
         year = 2024,
        month = jun,
       volume = {531},
       number = {2},
        pages = {2864-2880},
          doi = {10.1093/mnras/stae1318},
archivePrefix = {arXiv},
       eprint = {2401.15769},
 primaryClass = {astro-ph.GA},
       adsurl = {https://ui.adsabs.harvard.edu/abs/2024MNRAS.531.2864M}
}

@ARTICLE{Tanvir-2024MNRAS.527.7306T,
       author = {{Tanvir}, Tabassum S. and {Krumholz}, Mark R.},
        title = "{The metallicity dependence of the stellar initial mass function}",
      journal = {\mnras},
         year = 2024,
        month = jan,
       volume = {527},
       number = {3},
        pages = {7306-7316},
          doi = {10.1093/mnras/stad3581},
archivePrefix = {arXiv},
       eprint = {2305.20039},
 primaryClass = {astro-ph.GA},
       adsurl = {https://ui.adsabs.harvard.edu/abs/2024MNRAS.527.7306T}
}

@ARTICLE{Cheng-2023MNRAS.526.4004C,
       author = {{Cheng}, Chloe M. and {Villaume}, Alexa and {Balogh}, Michael L. and {Brodie}, Jean P. and {Mart{\'\i}n-Navarro}, Ignacio and {Romanowsky}, Aaron J. and {van Dokkum}, Pieter G.},
        title = "{Initial mass function variability from the integrated light of diverse stellar systems}",
      journal = {\mnras},
         year = 2023,
        month = dec,
       volume = {526},
       number = {3},
        pages = {4004-4023},
          doi = {10.1093/mnras/stad2967},
archivePrefix = {arXiv},
       eprint = {2309.14415},
 primaryClass = {astro-ph.GA},
       adsurl = {https://ui.adsabs.harvard.edu/abs/2023MNRAS.526.4004C}
}

@ARTICLE{Lee-2009ApJ...706..599L,
       author = {{Lee}, Janice C. and {Gil de Paz}, Armando and {Tremonti}, Christy and {Kennicutt}, Robert C., Jr. and {Salim}, Samir and {Bothwell}, Matthew and {Calzetti}, Daniela and {Dalcanton}, Julianne and {Dale}, Daniel and {Engelbracht}, Chad and {Funes}, S.~J. Jos{\'e} G. and {Johnson}, Benjamin and {Sakai}, Shoko and {Skillman}, Evan and {van Zee}, Liese and {Walter}, Fabian and {Weisz}, Daniel},
        title = "{Comparison of H{\ensuremath{\alpha}} and UV Star Formation Rates in the Local Volume: Systematic Discrepancies for Dwarf Galaxies}",
      journal = {\apj},
         year = 2009,
        month = nov,
       volume = {706},
       number = {1},
        pages = {599-613},
          doi = {10.1088/0004-637X/706/1/599},
archivePrefix = {arXiv},
       eprint = {0909.5205},
 primaryClass = {astro-ph.CO},
       adsurl = {https://ui.adsabs.harvard.edu/abs/2009ApJ...706..599L}
}

@ARTICLE{Zhou-2019MNRAS.485.5256Z,
       author = {{Zhou}, Shuang and {Mo}, H.~J. and {Li}, Cheng and {Zheng}, Zheng and {Li}, Niu and {Du}, Cheng and {Mao}, Shude and {Parikh}, Taniya and {Lane}, Richard R. and {Thomas}, Daniel},
        title = "{SDSS-IV MaNGA: stellar initial mass function variation inferred from Bayesian analysis of the integral field spectroscopy of early-type galaxies}",
      journal = {\mnras},
         year = 2019,
        month = jun,
       volume = {485},
       number = {4},
        pages = {5256-5275},
          doi = {10.1093/mnras/stz764},
archivePrefix = {arXiv},
       eprint = {1811.09799},
 primaryClass = {astro-ph.GA},
       adsurl = {https://ui.adsabs.harvard.edu/abs/2019MNRAS.485.5256Z}
}

@ARTICLE{Niu-2023ApJ...950..104N,
       author = {{Niu}, Zexi and {Yuan}, Haibo and {Liu}, Jifeng},
        title = "{Internal Calibration of LAMOST and Gaia DR3 GSP-Spec Stellar Abundances}",
      journal = {\apj},
         year = 2023,
        month = jun,
       volume = {950},
       number = {2},
          eid = {104},
        pages = {104},
          doi = {10.3847/1538-4357/accf8b},
archivePrefix = {arXiv},
       eprint = {2304.13951},
 primaryClass = {astro-ph.SR},
       adsurl = {https://ui.adsabs.harvard.edu/abs/2023ApJ...950..104N}
}

@ARTICLE{Siegel-2002ApJ...578..151S,
       author = {{Siegel}, M.~H. and {Majewski}, S.~R. and {Reid}, I.~N. and {Thompson}, I.~B.},
        title = "{Star Counts Redivivus. IV. Density Laws through Photometric Parallaxes}",
      journal = {\apj},
         year = 2002,
        month = oct,
       volume = {578},
       number = {1},
        pages = {151-175},
          doi = {10.1086/342469},
archivePrefix = {arXiv},
       eprint = {astro-ph/0206323},
 primaryClass = {astro-ph},
       adsurl = {https://ui.adsabs.harvard.edu/abs/2002ApJ...578..151S}
}

@ARTICLE{Bailer-2021AJ....161..147B,
       author = {{Bailer-Jones}, C.~A.~L. and {Rybizki}, J. and {Fouesneau}, M. and {Demleitner}, M. and {Andrae}, R.},
        title = "{Estimating Distances from Parallaxes. V. Geometric and Photogeometric Distances to 1.47 Billion Stars in Gaia Early Data Release 3}",
      journal = {\aj},
         year = 2021,
        month = mar,
       volume = {161},
       number = {3},
          eid = {147},
        pages = {147},
          doi = {10.3847/1538-3881/abd806},
archivePrefix = {arXiv},
       eprint = {2012.05220},
 primaryClass = {astro-ph.SR},
       adsurl = {https://ui.adsabs.harvard.edu/abs/2021AJ....161..147B}
}

@ARTICLE{Scalo-1986FCPh...11....1S,
       author = {{Scalo}, J.~M.},
        title = "{The Stellar Initial Mass Function}",
      journal = {\fcp},
         year = 1986,
        month = may,
       volume = {11},
        pages = {1-278},
       adsurl = {https://ui.adsabs.harvard.edu/abs/1986FCPh...11....1S}
}

@INPROCEEDINGS{Cignoni-2002ASPC..274..408C,
       author = {{Cignoni}, M. and {Castellani}, V. and {degl'Innocenti}, S. and {Petroni}, S. and {Prada Moroni}, P.~G.},
        title = "{The stellar content of the Galaxy: the white dwarf population}",
    booktitle = {Observed HR Diagrams and Stellar Evolution},
         year = 2002,
       editor = {{Lejeune}, Thibault and {Fernandes}, Jo{\~a}o},
       series = {Astronomical Society of the Pacific Conference Series},
       volume = {274},
        month = jan,
        pages = {408},
          doi = {10.48550/arXiv.astro-ph/0112289},
archivePrefix = {arXiv},
       eprint = {astro-ph/0112289},
 primaryClass = {astro-ph},
       adsurl = {https://ui.adsabs.harvard.edu/abs/2002ASPC..274..408C}
}

@ARTICLE{Chabrier-2003ApJ...586L.133C,
       author = {{Chabrier}, Gilles},
        title = "{The Galactic Disk Mass Function: Reconciliation of the Hubble Space Telescope and Nearby Determinations}",
      journal = {\apjl},
         year = 2003,
        month = apr,
       volume = {586},
       number = {2},
        pages = {L133-L136},
          doi = {10.1086/374879},
archivePrefix = {arXiv},
       eprint = {astro-ph/0302511},
 primaryClass = {astro-ph},
       adsurl = {https://ui.adsabs.harvard.edu/abs/2003ApJ...586L.133C}
}

@ARTICLE{Cappellari-2013MNRAS.432.1862C,
       author = {{Cappellari}, Michele and {McDermid}, Richard M. and {Alatalo}, Katherine and {Blitz}, Leo and {Bois}, Maxime and {Bournaud}, Fr{\'e}d{\'e}ric and {Bureau}, M. and {Crocker}, Alison F. and {Davies}, Roger L. and {Davis}, Timothy A. and {de Zeeuw}, P.~T. and {Duc}, Pierre-Alain and {Emsellem}, Eric and {Khochfar}, Sadegh and {Krajnovi{\'c}}, Davor and {Kuntschner}, Harald and {Morganti}, Raffaella and {Naab}, Thorsten and {Oosterloo}, Tom and {Sarzi}, Marc and {Scott}, Nicholas and {Serra}, Paolo and {Weijmans}, Anne-Marie and {Young}, Lisa M.},
        title = "{The ATLAS$^{3D}$ project - XX. Mass-size and mass-{\ensuremath{\sigma}} distributions of early-type galaxies: bulge fraction drives kinematics, mass-to-light ratio, molecular gas fraction and stellar initial mass function}",
      journal = {\mnras},
         year = 2013,
        month = jul,
       volume = {432},
       number = {3},
        pages = {1862-1893},
          doi = {10.1093/mnras/stt644},
archivePrefix = {arXiv},
       eprint = {1208.3523},
 primaryClass = {astro-ph.CO},
       adsurl = {https://ui.adsabs.harvard.edu/abs/2013MNRAS.432.1862C}
}

@ARTICLE{Almeida-2023,
       author = {{Almeida}, Andr{\'e}s and {Anderson}, Scott F. and {Argudo-Fern{\'a}ndez}, Maria and {Badenes}, Carles and {Barger}, Kat and {Barrera-Ballesteros}, Jorge K. and {Bender}, Chad F. and {Benitez}, Erika and {Besser}, Felipe and {Bizyaev}, Dmitry and {Blanton}, Michael R. and {Bochanski}, John and {Bovy}, Jo and {Brandt}, William Nielsen and {Brownstein}, Joel R. and {Buchner}, Johannes and {Bulbul}, Esra and {Burchett}, Joseph N. and {Cano D{\'\i}az}, Mariana and {Carlberg}, Joleen K. and {Casey}, Andrew R. and {Chandra}, Vedant and {Cherinka}, Brian and {Chiappini}, Cristina and {Coker}, Abigail A. and {Comparat}, Johan and {Conroy}, Charlie and {Contardo}, Gabriella and {Cortes}, Arlin and {Covey}, Kevin and {Crane}, Jeffrey D. and {Cunha}, Katia and {Dabbieri}, Collin and {Davidson}, James W., Jr. and {Davis}, Megan C. and {De Lee}, Nathan and {M{\'e}ndez Delgado}, Jos{\'e} Eduardo and {Demasi}, Sebastian and {Di Mille}, Francesco and {Donor}, John and {Dow}, Peter and {Dwelly}, Tom and {Eracleous}, Mike and {Eriksen}, Jamey and {Fan}, Xiaohui and {Farr}, Emily and {Frederick}, Sara and {Fries}, Logan and {Frinchaboy}, Peter and {Gaensicke}, Boris T. and {Ge}, Junqiang and {Gonz{\'a}lez {\'A}vila}, Consuelo and {Grabowski}, Katie and {Grier}, Catherine and {Guiglion}, Guillaume and {Gupta}, Pramod and {Hall}, Patrick and {Hawkins}, Keith and {Hayes}, Christian R. and {Hermes}, J.~J. and {Hern{\'a}ndez-Garc{\'\i}a}, Lorena and {Hogg}, David W. and {Holtzman}, Jon A. and {Ibarra-Medel}, Hector Javier and {Ji}, Alexander and {Jofre}, Paula and {Johnson}, Jennifer A. and {Jones}, Amy M. and {Kinemuchi}, Karen and {Kluge}, Matthias and {Koekemoer}, Anton and {Kollmeier}, Juna A. and {Kounkel}, Marina and {Krishnarao}, Dhanesh and {Krumpe}, Mirko and {Lacerna}, Ivan and {Jakson Assuncao Lago}, Paulo and {Laporte}, Chervin and {Liu}, Ang and {Liu}, Chao and {Liu}, Xin and {Lopes}, Alexandre Roman and {Macktoobian}, Matin and {Malanushenko}, Viktor and {Maoz}, Dan and {Masseron}, Thomas and {Masters}, Karen L. and {Matijevic}, Gal and {McBride}, Aidan and {Medan}, Ilija and {Merloni}, Andrea and {Morrison}, Sean and {Myers}, Natalie and {M{\'e}sz{\'a}ros}, Szabolcs and {Negrete}, C. Alenka and {Nidever}, David L. and {Nitschelm}, Christian and {Oravetz}, Audrey and {Oravetz}, Daniel and {Pan}, Kaike and {Peng}, Yingjie and {Pinsonneault}, Marc H. and {Pogge}, Rick and {Qiu}, Dan and {Queiroz}, Anna Barbara de Andrade and {Ramirez}, Solange V. and {Rix}, Hans-Walter and {Fern{\'a}ndez Rosso}, Daniela and {Runnoe}, Jessie and {Salvato}, Mara and {Sanchez}, Sebastian F. and {Santana}, Felipe A. and {Saydjari}, Andrew and {Sayres}, Conor and {Schlaufman}, Kevin C. and {Schneider}, Donald P. and {Schwope}, Axel and {Serna}, Javier and {Shen}, Yue and {Sobeck}, Jennifer and {Song}, Ying-Yi and {Souto}, Diogo and {Spoo}, Taylor and {Stassun}, Keivan G. and {Steinmetz}, Matthias and {Straumit}, Ilya and {Stringfellow}, Guy and {S{\'a}nchez-Gallego}, Jos{\'e} and {Taghizadeh-Popp}, Manuchehr and {Tayar}, Jamie and {Thakar}, Ani and {Tissera}, Patricia B. and {Tkachenko}, Andrew and {Hernandez Toledo}, Hector and {Trakhtenbrot}, Benny and {Fernandez Trincado}, Jose G. and {Troup}, Nicholas and {Trump}, Jonathan R. and {Tuttle}, Sarah and {Ulloa}, Natalie and {Vazquez-Mata}, Jose Antonio and {Alfaro}, Pablo Vera and {Villanova}, Sandro and {Wachter}, Stefanie and {Weijmans}, Anne-Marie and {Wheeler}, Adam and {Wilson}, John and {Wojno}, Leigh and {Wolf}, Julien and {Xue}, Xiang-Xiang and {Ybarra}, Jason E. and {Zari}, Eleonora and {Zasowski}, Gail},
        title = "{The Eighteenth Data Release of the Sloan Digital Sky Surveys: Targeting and First Spectra from SDSS-V}",
      journal = {arXiv e-prints},
         year = 2023,
        month = jan,
          eid = {arXiv:2301.07688},
        pages = {arXiv:2301.07688},
          doi = {10.48550/arXiv.2301.07688},
archivePrefix = {arXiv},
       eprint = {2301.07688},
 primaryClass = {astro-ph.GA},
       adsurl = {https://ui.adsabs.harvard.edu/abs/2023arXiv230107688A}
}

@PHDTHESIS{Best-2018PhDT.......159B,
       author = {{Best}, William M.~J.},
        title = "{Ultracool Demography with a Volume-Limited Census of the Solar Neighborhood}",
       school = {University of Hawaii, Manoa},
         year = 2018,
        month = jan,
       adsurl = {https://ui.adsabs.harvard.edu/abs/2018PhDT.......159B}
}

@ARTICLE{Wu-2011A&A...525A..71W,
       author = {{Wu}, Yue and {Singh}, H.~P. and {Prugniel}, P. and {Gupta}, R. and {Koleva}, M.},
        title = "{Coud{\'e}-feed stellar spectral library - atmospheric parameters}",
      journal = {\aap},
         year = 2011,
        month = jan,
       volume = {525},
          eid = {A71},
        pages = {A71},
          doi = {10.1051/0004-6361/201015014},
archivePrefix = {arXiv},
       eprint = {1009.1491},
 primaryClass = {astro-ph.SR},
       adsurl = {https://ui.adsabs.harvard.edu/abs/2011A&A...525A..71W}
}

@ARTICLE{Du-2021RAA....21..202D,
       author = {{Du}, Bing and {Luo}, A. -Li and {Zhang}, Shuo and {Kong}, Xiao and {Guo}, Yan-Xin and {Li}, Yin-Bi and {Zuo}, Fang and {Wang}, You-Fen and {Chen}, Jian-Jun and {Zhao}, Yong-Heng},
        title = "{LASPM: the LAMOST stellar parameter pipeline for M-type stars and application to the sixth and seventh data release (DR6 and DR7)}",
      journal = {Research in Astronomy and Astrophysics},
         year = 2021,
        month = oct,
       volume = {21},
       number = {8},
          eid = {202},
        pages = {202},
          doi = {10.1088/1674-4527/21/8/202},
       adsurl = {https://ui.adsabs.harvard.edu/abs/2021RAA....21..202D}
}

@ARTICLE{Souto-2022ApJ...927..123S,
       author = {{Souto}, Diogo and {Cunha}, Katia and {Smith}, Verne V. and {Allende Prieto}, C. and {Covey}, Kevin and {Garc{\'\i}a-Hern{\'a}ndez}, D.~A. and {Holtzman}, Jon A. and {J{\"o}nsson}, Henrik and {Mahadevan}, Suvrath and {Majewski}, Steven R. and {Masseron}, Thomas and {Pinsonneault}, Marc and {Schneider}, Donald P. and {Shetrone}, Matthew and {Stassun}, Keivan G. and {Terrien}, Ryan and {Zamora}, Olga and {Stringfellow}, Guy S. and {Lane}, Richard R. and {Nitschelm}, Christian and {Rojas-Ayala}, B{\'a}rbara},
        title = "{Detailed Chemical Abundances for a Benchmark Sample of M Dwarfs from the APOGEE Survey}",
      journal = {\apj},
         year = 2022,
        month = mar,
       volume = {927},
       number = {1},
          eid = {123},
        pages = {123},
          doi = {10.3847/1538-4357/ac4891},
archivePrefix = {arXiv},
       eprint = {2201.00891},
 primaryClass = {astro-ph.SR},
       adsurl = {https://ui.adsabs.harvard.edu/abs/2022ApJ...927..123S}
}

@ARTICLE{J-2020AJ....160..120J,
       author = {{J{\"o}nsson}, Henrik and {Holtzman}, Jon A. and {Allende Prieto}, Carlos and {Cunha}, Katia and {Garc{\'\i}a-Hern{\'a}ndez}, D.~A. and {Hasselquist}, Sten and {Masseron}, Thomas and {Osorio}, Yeisson and {Shetrone}, Matthew and {Smith}, Verne and {Stringfellow}, Guy S. and {Bizyaev}, Dmitry and {Edvardsson}, Bengt and {Majewski}, Steven R. and {M{\'e}sz{\'a}ros}, Szabolcs and {Souto}, Diogo and {Zamora}, Olga and {Beaton}, Rachael L. and {Bovy}, Jo and {Donor}, John and {Pinsonneault}, Marc H. and {Poovelil}, Vijith Jacob and {Sobeck}, Jennifer},
        title = "{APOGEE Data and Spectral Analysis from SDSS Data Release 16: Seven Years of Observations Including First Results from APOGEE-South}",
      journal = {\aj},
         year = 2020,
        month = sep,
       volume = {160},
       number = {3},
          eid = {120},
        pages = {120},
          doi = {10.3847/1538-3881/aba592},
archivePrefix = {arXiv},
       eprint = {2007.05537},
 primaryClass = {astro-ph.GA},
       adsurl = {https://ui.adsabs.harvard.edu/abs/2020AJ....160..120J}
}

@INPROCEEDINGS{Kroupa-2008mru..conf..227K,
       author = {{Kroupa}, P.},
        title = "{The stellar initial mass function of metal-rich populations}",
    booktitle = {The Metal-Rich Universe},
         year = 2008,
       editor = {{Israelian}, Garik and {Meynet}, Georges},
        month = jan,
        pages = {227},
          doi = {10.48550/arXiv.astro-ph/0703282},
archivePrefix = {arXiv},
       eprint = {astro-ph/0703282},
 primaryClass = {astro-ph},
       adsurl = {https://ui.adsabs.harvard.edu/abs/2008mru..conf..227K}
}

@INPROCEEDINGS{Li-2006AAS...20919803L,
       author = {{Li}, Di and {Guan}, X. and {Dai}, Y.},
        title = "{What is the True Core Mass Function?}",
    booktitle = {American Astronomical Society Meeting Abstracts},
         year = 2006,
       series = {American Astronomical Society Meeting Abstracts},
       volume = {209},
        month = dec,
          eid = {198.03},
        pages = {198.03},
       adsurl = {https://ui.adsabs.harvard.edu/abs/2006AAS...20919803L}
}

@ARTICLE{Lee-2020SSRv..216...70L,
       author = {{Lee}, Yueh-Ning and {Offner}, Stella S.~R. and {Hennebelle}, Patrick and {Andr{\'e}}, Philippe and {Zinnecker}, Hans and {Ballesteros-Paredes}, Javier and {Inutsuka}, Shu-ichiro and {Kruijssen}, J.~M. Diederik},
        title = "{The Origin of the Stellar Mass Distribution and Multiplicity}",
      journal = {\ssr},
         year = 2020,
        month = jun,
       volume = {216},
       number = {4},
          eid = {70},
        pages = {70},
          doi = {10.1007/s11214-020-00699-2},
archivePrefix = {arXiv},
       eprint = {2006.05778},
 primaryClass = {astro-ph.GA},
       adsurl = {https://ui.adsabs.harvard.edu/abs/2020SSRv..216...70L}
}

@PROCEEDINGS{Cor-2005ASSL..327.....C,
        title = "{The Initial Mass Function 50 years later}",
    booktitle = {The Initial Mass Function 50 Years Later},
         year = 2005,
       editor = {{Corbelli}, E. and {Palla}, F. and {Zinnecker}, H.},
       series = {Astrophysics and Space Science Library},
       volume = {327},
        month = jan,
          doi = {10.1007/978-1-4020-3407-7},
       adsurl = {https://ui.adsabs.harvard.edu/abs/2005ASSL..327.....C}
}

@ARTICLE{Cappellari-2012Natur.484..485C,
       author = {{Cappellari}, Michele and {McDermid}, Richard M. and {Alatalo}, Katherine and {Blitz}, Leo and {Bois}, Maxime and {Bournaud}, Fr{\'e}d{\'e}ric and {Bureau}, M. and {Crocker}, Alison F. and {Davies}, Roger L. and {Davis}, Timothy A. and {de Zeeuw}, P.~T. and {Duc}, Pierre-Alain and {Emsellem}, Eric and {Khochfar}, Sadegh and {Krajnovi{\'c}}, Davor and {Kuntschner}, Harald and {Lablanche}, Pierre-Yves and {Morganti}, Raffaella and {Naab}, Thorsten and {Oosterloo}, Tom and {Sarzi}, Marc and {Scott}, Nicholas and {Serra}, Paolo and {Weijmans}, Anne-Marie and {Young}, Lisa M.},
        title = "{Systematic variation of the stellar initial mass function in early-type galaxies}",
      journal = {\nat},
         year = 2012,
        month = apr,
       volume = {484},
       number = {7395},
        pages = {485-488},
          doi = {10.1038/nature10972},
archivePrefix = {arXiv},
       eprint = {1202.3308},
 primaryClass = {astro-ph.CO},
       adsurl = {https://ui.adsabs.harvard.edu/abs/2012Natur.484..485C}
}








\bsp	
\label{lastpage}
\end{document}